\documentclass[a4paper,onecolumn,superscriptaddress,10pt, accepted=2019-02-26]{quantumarticle}
\pdfoutput=1
\usepackage[numbers, sort&compress, merge]{natbib}
\usepackage[utf8]{inputenc}
\usepackage[english]{babel}
\usepackage[mathscr]{euscript}
\usepackage[T1]{fontenc}
\usepackage{bigints}
\usepackage{amsmath}
\allowdisplaybreaks
\usepackage{hyperref}
\hypersetup{
    colorlinks=true,
    linkcolor=blue,
    filecolor=magenta,      
    urlcolor=cyan,
}
\usepackage{graphicx}  
\usepackage{dcolumn}  
\usepackage{bm}          
\usepackage{amssymb} 
\usepackage{comment}
\usepackage{amsthm}
\usepackage{booktabs}
\usepackage{epstopdf}
\usepackage{float}
\usepackage{thmtools}
\usepackage{thm-restate}
\usepackage[capitalise]{cleveref} 

\restylefloat{table}
\newcommand{\be}{\begin{eqnarray}}
\newcommand{\ee}{\end{eqnarray} }
\newcommand{\benn}{\begin{eqnarray*}}
\newcommand{\eenn}{\end{eqnarray*}}
\newcommand{\txt}{\textrm}
\newcommand{\X}{\rangle\!\langle}  

\newcommand{\D}{\txt{d}}              
\newcommand{\bpr}{\begin{proof}}
\newcommand{\epr}{\end{proof}}   
\newtheorem{theorem}{Theorem}[section]

\newtheorem*{theorem*}{Theorem}
\newtheorem{corollary}{Corollary}[theorem]
\newtheorem*{corollary*}{Corollary}
\newtheorem{definition}{Definition}[section]
\newtheorem{eg}{Example}
\newtheorem*{egcontd}{Example 1 (continued)}
\newtheorem*{eg3contd}{Example 3 (continued)}
\newtheorem{lemma}{Lemma}[section]
\newtheorem{prop}{Proposition}[section]
\newtheorem*{prop*}{Proposition}
\DeclareMathOperator{\Tr}{tr}
\DeclareMathOperator*{\argmax}{arg\,max}

\begin{document}
\title{Minimax quantum state estimation under Bregman divergence}
\author{Maria Quadeer}
\email{mariaquadeer@gmail.com}
\author{Marco Tomamichel}
\author{Christopher Ferrie}
\affiliation{Centre for Quantum Software and Information,
		   University of Technology Sydney,
                Ultimo NSW 2007, Australia}
\begin{abstract}
We investigate minimax estimators for quantum state tomography under general Bregman divergences. First, generalizing the work of Koyama et al. [\href{http://www.mdpi.com/1099-4300/19/11/618}{Entropy 19, 618 (2017)}] for relative entropy, we find that given any estimator for a quantum state, there always exists a sequence of Bayes estimators that asymptotically perform at least as well as the given estimator, on any state. Second, we show that there always exists a sequence of priors for which the corresponding sequence of Bayes estimators is asymptotically minimax (i.e.~it minimizes the worst-case risk). Third, by re-formulating Holevo's theorem for the covariant state estimation problem in terms of estimators, we find that there exists a covariant measurement that is, in fact, minimax (i.e.~it minimizes the worst-case risk). Moreover, we find that a measurement that is covariant only under a unitary 2-design is also minimax. Lastly, in an attempt to understand the problem of finding minimax measurements for general state estimation, we study the qubit case in detail and find that every spherical 2-design is a minimax measurement.
\end{abstract}

\maketitle 
\section{Introduction}
Quantum state tomography~\cite{Fano1957,Pauli58} refers to the process of determining an unknown quantum state of a physical system by performing quantum measurements. Any information processing task necessarily involves verifying the output of a quantum channel which mandates the study of quantum state tomography, apart from the unavoidable theoretical necessity. With recent developments leading to a transition of quantum computation from theory to practice, the verification of quantum systems and processes is of particular importance. 

Given an unknown quantum state with no prior knowledge, it is clear that the measurement must be \textit{informationally complete}, i.e. a measurement with outcome statistics sufficient to fully specify the quantum state~\cite{Watrous_2018}. Conventional data-processing techniques such as \textit{direct inversion} and \textit{maximum likelihood estimation}, thus, implicitly assume that the measurement statistics are informationally complete.

In direct inversion, given a fixed informationally complete measurement, one identifies the frequencies of each outcome with the corresponding probabilities. Then, by inverting Born's rule one obtains a unique \textit{estimator} for the density operator that reproduces the measurement statistics; an estimator is defined as a map on the set of measurement outcomes $\mathcal{X}$, $\hat{\rho}:\mathcal{X}\mapsto\mathcal{S}(\mathcal{H})$, where $\mathcal{S}(\mathcal{H})$ is the set of density operators on the underlying Hilbert space $\mathcal{H}$ that describes the physical system. However, this strategy suffers from the drawback that such an estimator might not be a \textit{physical} state and would yield negative eigenvalues. 
\begin{eg}\label{eg1}
Suppose one measures an unknown quantum state in $\mathbb{C}^2$ along the x, y and z directions. Assuming that each of the measurements are performed only once, let us suppose that each of the outcome is `up'. Thus, $n_x=n_y=n_z=1$ and $N_x=N_y=N_z=1$, so that $p_x=n_x/N_x=1$, etc. Now, an estimator that would yield the same probabilities would be the one with the Bloch vector: $(2p_x-1,2p_y-1,2p_z-1)=(1,1,1)$. This is an invalid quantum state as it lies outside the Bloch ball, and thus necessarily has negative eigenvalues.
\end{eg}

Reference \cite{Hradil97} referred to such a shortcoming of direct inversion and proposed an alternative that enforces positivity on the estimator, called \textit{maximum likelihood estimation}.

A likelihood functional $\mathcal{L}[\rho]:\mathcal{S}(\mathcal{H})\mapsto [0,1] $ is the probability of observing a data set $\mathbb{D}$ given that the system is in the state $\rho$~:
\be\label{eq1}
\mathcal{L}[\rho]=p(\mathbb{D}|\rho)
\ee
The data set $\mathbb{D}$ is characterized by the outcome set of the given measurement $\{E_1,..., E_N|E_i\geq 0, \sum_{i=1}^N E_i=\mathbb{I}\}$. Thus, $p(\mathbb{D}|\rho)=\prod_{i=1}^N(\txt{Tr}[E_i\rho])^{n_i}$ where $n_i$ is the number of times the $i$-th outcome is recorded in $\mathbb{D}$. Maximum likelihood estimation involves maximizing Equation \eqref{eq1} over the space of density operators $\mathcal{S}(\mathcal{H})$, and thus obtaining as the estimator the state that maximizes the likelihood functional. 

The problem with MLE is that the estimator $\hat{\rho}_{MLE}$ can be rank-deficient. A rank-deficient estimator is not good, as it would mean that by performing only finite number of measurements we are absolutely certain to rule out many possibilities. This kind of certainty must be bogus, suggesting that there has to be a better estimator. Let us look at \cref{eg1} once again to illustrate this point.
\begin{egcontd}
Given the choice of measurement and the corresponding outcomes, the likelihood functional is $\mathcal{L}[\rho]=(1 + r_x) (1 + r_y) (1 + r_z) / 6^3 $, which needs to be maximized under the constraint $\|\vec{r}\|\leq1$---that characterizes the physical set of states in $\mathbb{C}^2$. This implies that $r_x=r_y=r_z=1/\sqrt{3}$, which corresponds to an estimator that is a pure state.
\end{egcontd}
This shows that state estimators that are unphysical in direct inversion get mapped to the closest physical states in MLE, that lie on the state space boundary, and are thus rank-deficient. In fact, it can be shown~\cite{Blume-Kohout2010} that if there exists a $\rho_{DI}$ obtained via direct inversion over a data set $\mathbb{D}$ which is physical, then, it also maximizes the likelihood functional, i.e. $\hat{\rho}_{DI}=\hat{\rho}_{MLE}$. Although, \cref{eg1} is an instance of an extreme case where probabilities are approximated by frequencies of a single measurement, it suffices to illustrate that in direct inversion as well as MLE, all that one cares about is to obtain an estimate of the true state that reproduces the observed measurement statistics, regardless of the fact that in the light of new data the state's estimate might change completely. Reference \cite{Blume-Kohout2010} gives a detailed critique of both direct inversion and MLE, proposing  \textit{Bayesian Mean Estimation} (BME) to be a more plausible estimation technique. Moreover, it has been shown that such an estimation technique is quantitatively better in reference \cite{Chris&Blume-Kohout2018}.

Generally speaking, in estimation theory~\cite{Lehmman98}, the average measure of closeness of an estimator to the actual state is defined as the \textit{risk},
\be\label{riskdef}
R(\rho, \hat{\rho})=\mathbb{E}_{X|\rho}[L(\rho, \hat{\rho}(X))],
\ee
where $X$ is the random variable corresponding to the measurement outcomes and $L$ is a distance-measure between the true state and the estimator. One way of choosing an optimal estimator is to look at the \emph{average risk}---defined as the expectation of risk with respect to a \emph{prior} distribution over $\mathcal{S}(\mathcal{H})$. Then, by minimizing the average risk over the set of all probability distributions over $\mathcal{S}(\mathcal{H})$, one obtains what is called a \emph{Bayes estimator}, $\hat{\rho}_B$~\cite[pg.~228]{Lehmman98}. In fact, it has been shown that the Bayes estimator is the mean if the loss function is the relative entropy \cite{Tanaka&Komaki2005}, while in reference \cite{Banerjee} the same was proved for a more general class of distance-measures called \emph{Bregman divergence} (see \cref{def2}), which generalizes two important distance-measures---relative entropy and Hilbert-Schmidt distance, but in the classical setting. We provide a proof for the quantum setting in \cref{A0} for completion. 

Now, the Bayesian mean estimator for a prior distribution $\pi(\rho)$ is given by
\be\label{eq3}
\hat{\rho}_{B}(\mathbb{D})=\int_{\mathcal{S}(\mathcal{H})} p(\rho|\mathbb{D})\rho\, \D\rho,
\ee
where $p(\rho|\mathbb{D})$ is the posterior probability density given by the Bayes rule:
\be\label{eq4}
p(\rho|\mathbb{D})=\frac{p(\mathbb{D}|\rho)\pi(\rho)}{p(\mathbb{D})},
\ee
and $p(\mathbb{D})=\int_{\mathcal{S}(\mathcal{H})} \D\pi(\rho)p(\mathbb{D}|\rho)$. However, BME can yield nonsensical estimators if one starts with a \emph{bad} prior, as the following example illustrates.
\begin{eg}
Consider a $\sigma_X$ measurement on an unknown quantum state $\rho$ in $\mathbb{C}^2$. Suppose there exists a prior $\pi(\rho)$ such that  it assigns zero measure to all states in $\mathbb{C}^2$ but $|-\X-|$. A single measurement outcome of `+' rules out the outcome `$-$' and thus annihilates the prior!
\end{eg}
In fact, it should be clear from the above example that some priors can be annihilated by a finite number of (independent) measurements. Thus, in general, one needs a \emph{robust}~\cite{Blume-Kohout2010} prior that cannot be annihilated in order to prevent rank-deficient estimates. However, the estimator's knowledge of the true state can \emph{still} be jeopardized in the presence of an adversary who provides her with a wrong prior. Therefore, although BME seems to be the best bet, it remains inherently ambiguous due to its dependence on the choice of priors. A systematic approach towards deriving \emph{optimality} criteria for priors is thus a compelling problem.

The \textit{minimax} approach, complementary to BME, seems to be doing just that. In \emph{classical statistics}, the problem of estimating probability distributions (analogous to state estimation) has been studied using the \textit{minimax approach}~\cite{CLARKE1994, Merhav&Feder1998, Xie&Barron2000, Komaki_Classical2010}, that offers an alternative characterization of optimality of estimators. In the minimax approach, one looks at the space of all possible estimators defined on $\mathcal{X}$ and, for each estimator $\hat{\rho}$, picks the state $\rho$ for which it has the worst performance or \textit{risk} (quantified in terms of a suitably chosen distance-measure between the estimator and the true state). Then, the \textit{minimax estimator} is the one that has the \emph{best} worst-case risk.  Such an estimator necessarily works for all states $\rho \in \mathcal{S}(\mathcal{H})$. It can be shown~\cite{CLARKE1994} that such a minimax estimator is a Bayes estimator given a particular choice of `non-informative' prior. Thus, the solution to the minimax problem leads to a natural identification of a prior. 

However, as pointed out in reference \cite{Blume-Kohout2010}, no such rigorous statements were known for the quantum analogue of the problem until then. Recently, the authors of reference \cite{Komaki2017} have studied the quantum minimax estimation problem in analogy to the classical problem \cite{Komaki_Classical2010}, quantifying the estimator's risk in terms of relative entropy. To summarize, they find that given an unknown quantum state $\rho$ and some estimator $\hat{\rho}$ of it, there always exists a sequence of Bayes estimators that perform at least as well as $\hat{\rho}$ in the limiting case. Moreover, they show that there always exists a class of priors, called \textit{latent information priors} (although, conventionally, such priors are called \textit{least favourable}, and we shall follow the convention!) for which there is a corresponding sequence of Bayes estimators whose limit is \textit{minimax}. Finally, they define a \textit{minimax POVM} as a POVM that minimizes the minimax risk, see \cref{def4}, and study the qubit ($\mathbb{C}^2$) case in detail, obtaining the class of the least favourable priors as well as the minimax POVM for $\mathbb{C}^2$.

This paper is divided into six sections---we discuss our main results in \cref{sec2}, followed by \cref{sec3a} that contains the formalism and \cref{sec3b} that contains the proofs in detail. Finally, in \cref{sec4}, we discuss the state estimation problem for $\mathbb{C}^2$. In \cref{sec5}, we summarize the results and outline future work.
\section{Main Results}\label{sec2}
Bayesian mean estimation is arguably a more plausible approach towards state estimation as opposed to maximum likelihood estimation or direct inversion. However, the performance of BME is tied to the choice of the prior. A complementary approach towards the state estimation problem---the \textit{minimax} approach provides a window to explore all possible classes of priors,  enabling one to narrow down those that are consistent with the requirements of both the Bayesian and the minimax analysis. 

We extend the work done in reference \cite{Komaki2017} on minimax analysis (as discussed earlier) to a more general class of distance-measures called the \textit{Bregman divergence}, see \cref{def2}, that generalizes both relative entropy and Hilbert-Schmidt distance. We also generalize the minimax POVM for $\mathbb{C}^2$ to Hilbert-Schmidt distance, finding that  such a minimax POVM is a spherical 2-design.  Moreover, by re-formulating Holevo's theorem~\cite[pg.~171]{holevo82} for the \textit{covariant state estimation problem} in terms of estimators, we find that a \textit{covariant} POVM is, infact, minimax with Bregman divergence as the distance-measure. Let us discuss these results in detail,  informally, postponing the formal statements and proofs to \cref{sec3b}.

\begin{restatable}{result}{thmone}\label{thm1}
For any estimator $\hat{\rho}$, there always exists a sequence of Bayes estimators such that the limit of the sequence performs at least as well as $\hat{\rho}$, i.e.
\benn
R(\rho,\hat{\rho})\geq R\Big(\rho,\lim_{n\to\infty}\hat{\rho}_B^{\pi_n}\Big),~~~~\forall \rho\in\mathcal{S}(\mathcal{H}).
\eenn
\end{restatable}
So, for a given quantum state and some estimator, the above result says that one can always find a sequence of Bayes estimators that asymptotically perform at least as well as the estimator. That there exists such a sequence of Bayes estimators means that there exists a corresponding convergent sequence of priors, see Equation \eqref{eq3}. However, in general, the Bayes estimator is uniquely defined only up to the null set of $p_\pi$ comprised of outcomes that have zero probability under $\pi$. This illustrates that one cannot replace any given estimator by a corresponding Bayes estimator. However, one can still find a sequence of Bayes estimators that, in the limiting case, perform at least as well as the given estimator. This is an interesting result as it benchmarks BME against any given estimation technique. A related question is: ``Does there exist a Bayes estimator that is also minimax?'' Well, if one is given a minimax estimator, then \cref{thm1} tells us that there exists a sequence of Bayes estimators that are at least as good as the minimax estimator, which in turns implies that the limit itself is minimax. In fact, the next result gives us a Bayesian procedure to arrive at such a minimax estimator.
\begin{restatable}{result}{thmtwo}\label{thm2}
There always exists a sequence of priors such that the limit of the sequence maximizes the average risk of a Bayes estimator,
\be\label{<risk>}
r(\pi,\hat{\rho}_B^{\pi})=\int_{\mathcal{S}(\mathcal{H})}\D\pi(\rho) R(\rho,\hat{\rho}_B^{\pi}),
\ee and the limit of the respective sequence of Bayes estimators minimizes the worst-case risk, i.e., it is minimax.
\end{restatable}
We find that the prior that maximizes the average risk of the Bayes estimator---referred to as the least favourable prior (as it maximizes the minimum possible average risk) is the limit of a convergent sequence of priors, such that the limit of the corresponding sequence of Bayes estimators is minimax. Note that the average risk with relative entropy as the loss function is the average of the maximal accessible information---Holevo information for the ensemble $\{\pi(\rho|x),\rho\}$ with respect to the total probability of measurement outcomes. So, maximizing this average Holevo information of the given ensemble over all possible priors is like asking ``What is the prior for which the posterior yields maximum accessible information for the ensemble $\{\pi(\rho|x),\rho\}$?'' However, no such interpretation can be made for general loss functions such as Bregman divergence.

At this point, it must be noted that the underlying measurement in all the analysis done so far is fixed. Thus, a least favourable prior is inherently tied to the measurement. Any construction of such a prior necessarily implies that we also need to find a class of measurements for which it works. Although it is not clear if one would be able to solve this problem in general, one can do so for at least a subset of the estimation problem---the \textit{covariant state estimation} problem. 

In covariant state estimation, given a fixed state $\rho_0$, one is interested in estimating the states $\rho_\theta$ such that $\rho_\theta\in\{V_g\rho_0V_g^\dagger\}$ where $g\in G$ is a group element acting on the parameter space $\Theta$ with $\theta\in\Theta$ and $V_g$ is the projective unitary representation of the parametric group $G$. By generalizing Holevo's theorem~\cite[Theorem 3.1]{holevo82} to Bregman divergences, we obtain a least favourable prior.
\begin{restatable}{lemma}{Covlemma}\label{lem2}
The uniform measure on the parameter space $\Theta$ is a least favourable prior for covariant measurements.
\end{restatable}
A covariant measurement is a measurement that reflects the transformation of the state under the group action appropriately in the outcome statistics (see \cref{def5}). Now, the question is: ``What is the measurement that minimizes the worst-case risk?'' An answer to this question is closely related to finding the class of measurements for the least favourable priors. The only additional information that is needed is if such a class of measurements also minimizes the average risk with respect to the least favourable prior. It turns out that this is indeed the case as far as covariant estimation is concerned.
\begin{restatable}{result}{thmthree}\label{thm3}
There exists a covariant measurement $\mathscr{P}_c$ which is minimax for covariant state estimation. Moreover, if there exists a measurement $\mathscr{P}^\prime_c$ which is covariant under a subgroup $H$ of $G$ such that $\{V_h|~h\in H\}$ forms a unitary 2-design,  where $V_h$ is the projective unitary representation of the subgroup $H$, and $\mathscr{P}_c$ and $\mathscr{P}^\prime_c$ have the same seed\footnote{See \cref{lemE3}.}, then $\mathscr{P}^\prime_c$ is also minimax.
\end{restatable}

It is not so straightforward to generalize these results to the general state estimation problem. To better understand the situation for the general case, we look at the simplest system of a single qubit (extending the results of reference \cite{Komaki2017} to Hilbert-Schmidt distance---details in \cref{sec5}. In particular, we find that every \emph{spherical 2-design} in $\mathbb{C}^2$ is a minimax POVM.

\section{Formalism}\label{sec3a}
Consider a quantum system $\mathcal{S}$ described by a finite-dimensional Hilbert space $\mathcal{H}$ with $\mathcal{S}(\mathcal{H})$ as  the set of density operators on $\mathcal{H}$. Then, consider a quantum measurement to be an experiment in which the quantum system $\mathcal{S}$ is measured and let $\mathcal{X}$ be the corresponding outcome space of the measurement outcomes. Each possible event of the experiment can be identified with a subset B $\subseteq\mathcal{X}$, the event being `the measurement outcome $x$ lies in B'. The probability distribution of the events is thus defined over a $\Sigma$-algebra of the measurable subsets B $\subseteq\mathcal{X}$. To be in touch with physical reality, we choose the outcome space to be a Haursdorff space, i.e. a topological space where for any $x_1,x_2\in\mathcal{X}$ there exist two disjoint open sets $X_1,X_2\subset\mathcal{X}$ such that $x_1\in X_1$ and $x_2\in X_2$. This ensures that the $\Sigma$-algebra is a Borel $\Sigma$-algebra generated by countable intersections, countable unions and relative complements of open subsets of $\mathcal{X}$. Let $\mathcal{P}(\mathcal{H})$ be the set of positive operators on $\mathcal{H}$. 
\begin{definition}[Quantum measurement]\label{def1}
A Positive Operator-Valued Measure (POVM) is a map $P:\Sigma\mapsto\mathcal{P}(\mathcal{H})$, where $\Sigma$ is the $\Sigma$-algebra of all measurable subsets of $\mathcal{X}$. Thus, a POVM associates an operator P(B) to each $B\in\Sigma$ satisfying the following:
\begin{enumerate}
\item $P(B)\geq0$,~~~~~~$\forall B\in\Sigma$.
\item $P(\mathcal{X})=\mathbb{I}$.
\item $P\Big(\bigcup\limits_{i=1}^\infty B_i\Big)=\sum\limits_{i=1}^\infty P(B_i)$, ~~~~~~where $\forall B_i,B_j~~\txt{s.t.}~B_i \cap B_j =\emptyset~$.
\end{enumerate}
\end{definition}
The set of all POVMs on $\Sigma$ forms a convex set denoted by $\mathscr{P}$. A  POVM  is \textit{informationally complete}~\cite{Watrous_2018} if the operators $\{P(B)\}$ span $L(\mathcal{H})$, the space of linear operators on $\mathcal{H}$. The measurement statistics of such an IC-POVM is sufficient to determine, uniquely, all possible states that the quantum system could be in, in the limit when an infinite number of measurements are performed. Optimization of data-processing deals with the practical aspect of not having infinite resources and minimizing the corresponding statistical error. Reference \cite{Bisio2017} reviews the theoretical development of optimization techniques in quantum tomography based on informationally complete measurements. However, in this paper, we make no assumptions on the POVM. In fact, we look at an alternative definition for an optimal POVM---to be discussed later in this section. 

The following lemma (see \cref{A1} for proof) provides a convenient way of representing a POVM as an operator-valued density.
\begin{restatable}[Existence of a POVM density]{lemma}{RadonNikodym}\label{lem0}
Every P $\in\mathscr{P}$ admits a density, i.e. for any POVM P there exists a finite measure $\mu(dx)$ over $\mathcal{X}$ such that $\mu(\mathcal{X})=1$ and
\be
P(B)=\int_B \D\mu(x) M(x),
\ee
with $M(x)\geq0$, and Tr$[M(x)]=d~\mu-$almost everywhere.
\end{restatable}

The conditional probability of the event `the measurement outcome $x'$ lies in $B$ given that the system is in a state $\rho$' is given by Born's rule as
\be\label{eq5}
\txt{Pr}[x'\in B|~\rho]=\Tr P(B)\rho=\int_B \D\mu(x) \Tr M(x)\rho,
\ee
or, in the differential form as 
\be\label{Born'sruleDiff}
\D p(x|\rho)=\D\mu(x) \Tr M(x)\rho
\ee
which will come in handy later. Now that we have defined a quantum measurement, we proceed with the formulation. 

In estimation theory~\cite{Lehmman98}, one typically parametrizes the system $\mathcal{S}$ by a parameter $\theta$. The data set of the measurement outcomes is represented by a \textit{random variable} X. Using this data set one estimates the parameter $\theta$ or more generally $\rho_\theta$---the \textit{estimand}. Succinctly, this involves two random variables $\Theta$ and X  defined as below:

\begin{itemize}
\item  In the Bayesian model, the quantity $\theta$ that parametrizes the system $\mathcal{S}$ is treated as a random variable $\Theta$. This random variable is defined over the parameter space $\Omega_\Theta$\footnote{However, we will abuse notation and refer to the parameter space as $\Theta$ from now onwards.} and is distributed according to an a-priori probability distribution $\pi_\Theta\in \mathcal{P}(\Theta)$ (where $\mathcal{P}(\Theta)$ is the set of all probability distributions on $\Theta$). 
\item X is the random variable associated with the  outcomes  of the measurement performed on the system $\mathcal{S}$, defined over the sample space $\mathcal{X}$. The outcomes of the measurement are conditioned on the random variable $\Theta$. Thus, X is distributed according to the conditional probability $p_X(x|\theta)$, given by Equation \eqref{eq5}.
\end{itemize}

The parameter space $\Theta$ is chosen to be a compact metric space. The set of all bounded continuous real-valued function on $\Theta$ is denoted by $\mathcal{C}(\Theta,\mathbb{R})$. The set of probability distributions $\mathcal{P}(\Theta)$ on $\Theta$ is endowed with a weak topology, which essentially defines the notion of \emph{weak convergence}.
\begin{definition}
A sequence of probability measures $\pi_n\in\mathcal{P}(\Theta)$ weak converges to $\mu$ if for
every $f\in\mathcal{C}(\Theta,\mathbb{R})$,
\be 
\int f \D\pi_n \rightarrow \int f \D\mu, ~~~\txt{as $n\rightarrow\infty$.}
\ee
\end{definition} 

Then, as $\Theta$ is a compact metric space and $\mathcal{P}(\Theta)$ is endowed with a weak topology, by \cite[Theorem 6.4]{KRP1967}, it implies that $\mathcal{P}(\Theta)$ is also a compact metric space.

The central problem in quantum state estimation is to obtain an \textit{estimator} of $\rho_\theta$. We define an \textit{estimator} as the map
\be\label{eq2}
\hat{\rho} : \mathcal{X} \mapsto \mathcal{S}(\mathcal{H}).
\ee
The value of $\hat{\rho}$(x) is the estimate of $\rho_\theta$ when the measurement outcome is $X=x$. We want $\hat{\rho}$(X) to be close to $\rho_\theta$, but $\hat{\rho}$(X) is a random variable. One way of defining a meaningful measure of closeness is by defining an expectation over the conditional distribution of X, Equation \eqref{eq5}. Let $L(\rho_\theta, \hat{\rho}(x))$ be the \textit{loss function} that quantifies the closeness of an estimated state $\hat{\rho}$(x) to the true state $\rho_\theta$. We assume two things about $L$:
\begin{enumerate}
\item $L(\rho_\theta, \hat{\rho}(x)) \geq 0$, $\forall\theta\in\Theta, ~\hat{\rho}$, with equality if and only if $\rho_\theta=\hat{\rho}(x)$.
\item $L(\rho_\theta,\rho_\theta)=0,~~~~\forall\theta\in\Theta$.
\end{enumerate}

The average measure of  closeness of $\hat{\rho}$(X) to $\rho_\theta$ is defined as the \textit{risk function}
\be\label{}
R(\rho_\theta, \hat{\rho})=\mathbb{E}_{X|\theta}[L(\rho_\theta, \hat{\rho}(X))].
\ee
One would like to obtain an estimator that minimizes the risk for all values of $\theta$. Obviously, this problem does not have a solution, i.e. there does not exist an estimator that uniformly minimizes the risk for all values of $\theta$ except for the case when $\rho_\theta$ is a constant. Instead, one can look at the following two quantities that are a good measure of the risk in a global sense:
\begin{enumerate}
\item Average risk:
\be 
r(\pi,\hat{\rho})=\int_{\Theta} \D\pi(\theta) R(\rho_\theta, \hat{\rho}),
\ee
where $\pi(\theta)$ is an a-priori distribution over the parameter space $\Theta$.
\item Worst-case/minimax risk:
\be\label{eq7}
\inf_{\hat{\rho}}\sup_{\theta} R(\rho_\theta, \hat{\rho}).
\ee 
\end{enumerate}
The estimator that minimizes the average risk is the Bayes estimator $\hat{\rho}_B$~\cite[pg.228]{Lehmman98}. In reference \cite{Tanaka&Komaki2005} it was shown that the Bayes estimator is the mean if the loss function is the relative entropy $D(\rho_\theta||\hat{\rho}(x))$, i.e.,
\be\label{eq13}
\hat{\rho}_B=x\mapsto\int_{\Theta} \D\pi(\theta|x) \rho_{\theta},
\ee 
where $\D\pi(\theta|x)$ is the posterior probability distribution  obtained via the Bayes rule, 
\benn
\D\pi(\theta|x)=\frac{\D p(x|\theta)}{\D p_\pi(x)}\D\pi(\theta),
\eenn
where $p_\pi(B)=\int_B \int_\Theta \D p(x|\theta)\D \pi(\theta)$. Note that in the continuous case, the likelihood ratio $\frac{p(x|\theta)}{p_\pi(x)}$ is replaced by the corresponding Radon-Nikodym derivative, which is defined uniquely upto the null set of $p_\pi$. In fact,  the same was proved \cite{Banerjee} for a more general class of distance-measures called \textit{Bregman divergence} which is the measure we will use in our analysis, but only in the classical setting.  We provide a proof for the quantum setting in \cref{A0} for completion. Let us now define Bregman divergence.
\begin{definition}[Bregman divergence for density matrices]\label{def2}
Let $f:[0,1]\mapsto \mathbb{R}$ be a strictly convex continuously-differentiable real-valued function. Then, the Bregman divergence between density matrices $\rho, \sigma$ is defined as
\benn
D_f(\rho,\sigma)=\Tr \big(f(\rho)-f(\sigma)-f'(\sigma)(\rho-\sigma)\big).
\eenn
\end{definition}
Bregman divergence generalizes two important classes of distance-measures: the relative entropy obtained by choosing $f:x\mapsto x\log x$ and the Hilbert-Schmidt distance (Schatten 2-norm) obtained by choosing $f:x\mapsto x^2$. 

Let us now look at a few of its important properties. First, Bregman divergence is invariant under unitary transformations of its arguments, i.e. $D_f(U\rho U^\dagger,U\sigma U^\dagger)=D_f(\rho,\sigma)$. Second, it is not a metric, as it is neither symmetric nor satisfies the triangle inequality, but by the strict convexity of f, $D_f(\rho,\sigma)\geq0$, with equality if and only if $\rho=\sigma$. Third, the convexity of f implies that $D_f(.,.)$ is convex in its first argument; it is jointly convex if $f''$ is operator convex and numerically non-increasing \cite{Virosztek2015}. Moreover, by generalizing the proof of lower semi-continuity of relative entropy as in reference \cite{Wehrl}, we obtain the lower semi-continuity of Bregman divergence (see \cref{lsc} for the proof). 

We are now ready to state and prove the main results of this paper.
\section{Proofs}\label{sec3b}
\subsection{Bayesian state estimation}
Formally, \cref{thm1} is stated as the following theorem.
\begin{theorem}
Let $\hat{\rho}:\mathcal{X}\mapsto\mathcal{S}(\mathcal{H})$ be an estimator. Then, there exists a convergent sequence of  priors such that the corresponding sequence of Bayes estimators $(\hat{\rho}_B^{\pi_n})_n$ converges, with
\be\label{eq9}
R(\rho_{\theta},\hat{\rho})\geq R\Big(\rho_{\theta},\lim_{n\to\infty}\hat{\rho}_B^{\pi_n}\Big),~~~~\forall \theta \in \Theta.
\ee 
\end{theorem}
\bpr
Consider the average distance between the Bayes estimator and a given estimator $\hat{\rho}$ for some prior $\pi\in\mathcal{P}(\Theta)$ as the map
\benn
g:\pi\mapsto D^f_{\hat{\rho}}(\pi)=\int_\mathcal{X}\D p_\pi(x)~D_f(\hat{\rho}_B^\pi(x),\hat{\rho}(x)).
\eenn
Now, the Bayes estimator is uniquely defined up to the null set of $p_\pi$. In fact, it is discontinuous on $\mathcal{X}$, see \cref{disctsBayes}, and the points of discontinuity belong to the null set of $p_\pi$. So, unless the null set of $p_\pi$ is empty, the Bayes estimator cannot be defined continuously since there can exist different sequences that converge to the same prior, but the limit of the corresponding sequences of Bayes estimators may not coincide on the null set of $p_\pi$. To deal with the discontinuity of the Bayes estimator, we consider closed subsets of $\mathcal{P}(\Theta)$ with the defining property that every element of these subsets renders the corresponding Bayes estimator continuous on  $\mathcal{X}$. Then, g is lower semi-continuous on each closed subset as Bregman divergence is lower-semi continuous (\cref{lsc}). Thus, there exists a prior $\pi_n$ in every subset that minimizes it on that subset. So, we look at the sequence of such priors $(\pi_n)_n$ and find that the corresponding sequence of Bayes estimators $(\hat{\rho}_B^{\pi_n})_n$ converges to a limit that has a risk lower than or equal to that of the given estimator. Let us now proceed with the proof.

We define the closed subsets of $\mathcal{P}(\Theta)$ as
\be\label{subset1}
\mathcal{P}_{\mu/n}=\Big\{\frac{\mu}{n}+\left(1-\frac{1}{n}\right)\pi\Big|\pi\in\mathcal{P}(\Theta)\Big\},
\ee
where $\mu$ is a measure such that $p_\mu(x)>0$, for all $x\in\mathcal{X}$. The latter condition ensures that the Bayes estimator for a prior that lies in $\mathcal{P}_{\mu/n}$ is continuous on $\mathcal{P}_{\mu/n}$. Then, as a closed subset of a compact set is compact, there exists a prior $\pi_n\in \mathcal{P}_{\mu/n}$ such that $D_{\hat{\rho}}^f(\pi_n)=\inf\limits_{\pi\in\mathcal{P}_{\mu/n}}D_{\hat{\rho}}^f(\pi)$. In fact, as $\mathcal{P}(\Theta)$ is a compact metric space, the sequence of priors $(\pi_n)_n$ has a convergent subsequence. Let us denote this subsequence as $(\pi^{'}_m)_m$. Let $n_m$ be such that $\pi_{n_m}=\pi^{'}_m$.

Then, the idea is to use the fact that each $\pi^{'}_m$ minimizes $D_{\hat{\rho}}^f$ on the corresponding closed subset $\mathcal{P}_{\mu/n_m}$ to obtain a suitable condition. To begin with, we define a prior in the neighbourhood of $\pi^{'}_{m+1}\in\mathcal{P}_{\mu/n_{m+1}}$ by taking a convex sum of it with another element in $\mathcal{P}_{\mu/n_{m+1}}$. Observe that $\Big(\frac{n_m}{n_{m+1}}\pi^{'}_{m+1}+(1-\frac{n_m}{n_{m+1}})\delta({\theta-\theta_0})\Big)$ lies in $\mathcal{P}_{\mu/n_{m+1}}$, for any $\theta_0\in\Theta$. So, we define a prior
\be\label{subset2}
\pi(\theta)=u\Bigg(\frac{n_m}{n_{m+1}}\pi^{'}_m+\Big(1-\frac{n_m}{n_{m+1}}\Big)\delta({\theta-\theta_0})\Bigg)+(1-u)\pi^{'}_{m+1},
\ee
with $0\leq u\leq1$. This is like considering a perturbation in the neighbourhood of $\pi^{'}_{m+1}$ and noting that the derivative of $D^f_{\hat{\rho}}(\pi)$ is positive as one approaches $\pi^{'}_{m+1}$ as it minimizes $D^f_{\hat{\rho}}(\pi)$  on the set $\mathcal{P}_{\mu/n_{m+1}}$. Thus, we have
\begin{flalign}\label{eqA12b}
0\leq\frac{\D}{\D u}D^f_{\hat{\rho}}(\pi)\Bigg|_{u=0}&=\frac{\D}{\D u}\int_\mathcal{X}\D p_\pi(x)~D_f(\hat{\rho}_B^\pi(x),\hat{\rho}(x))\Bigg|_{u=0}\notag\\
&=\int_\mathcal{X}\frac{\D p_{\pi}(x)}{\D u}\Bigg|_{u=0}D_f(\hat{\rho}_B^{\pi^{'}_{m+1}}(x),\hat{\rho}(x))+\int_\mathcal{X}\D p_{\pi^{'}_{m+1}}(x)\frac{\D }{\D u}D_f(\hat{\rho}_B^\pi(x),\hat{\rho}(x))\Bigg|_{u=0}.
\end{flalign}
Let $k=\frac{n_m}{n_{m+1}}$. Then, 
\benn
\D p_\pi(x)&=&\bigintss \D p(x|\theta)\Bigg(u\Big(k\pi^{'}_m(\theta)+(1-k)\delta({\theta-\theta_0})\Big)+(1-u)\pi^{'}_{m+1}(\theta)\Bigg) \D\theta.\\
\implies \frac{\D p_{\pi}(x)}{\D u}\Bigg|_{u=0}&=& k\D p_{\pi^{'}_m}(x)+(1-k)\D p(x|\theta_0)-\D p_{\pi^{'}_{m+1}}(x).
\eenn
So, the first term \eqref{eqA12b} is
\benn
\int_\mathcal{X}\Big(k\D p_{\pi^{'}_m}(x)+(1-k)\D p(x|\theta_0)-\D p_{\pi^{'}_{m+1}}(x)\Big)D_f\big(\hat{\rho}_B^{\pi^{'}_{m+1}}(x),\hat{\rho}(x)\big),
\eenn
while the second term, using \cref{lemC1}, is
\begin{flalign}\label{eqA13a}
&=\int_\mathcal{X}\D p_{\pi^{'}_{m+1}}(x)\frac{\D}{\D u}\Tr f\big(\hat{\rho}_B^\pi(x)\big)-f'\big(\hat{\rho}(x)\big)\big(\hat{\rho}_B^\pi(x)-\hat{\rho}(x)\big)\Bigg|_{u=0}\notag\\
&=\int_\mathcal{X}\D p_{\pi^{'}_{m+1}}(x)\Tr\Big(f'\big(\hat{\rho}_B^{\pi^{'}_{m+1}}(x)\big)-f'\big(\hat{\rho}(x)\big)\Big) \frac{\D}{\D u}\hat{\rho}_B^\pi(x)\Big|_{u=0}.
\end{flalign}

Let us now calculate the derivative of the Bayes estimator.
\begin{flalign}\label{eqA13}
\frac{\D}{\D u}\hat{\rho}_B^\pi(x)\Big|_{u=0}&=\frac{\D}{\D u}\bigintssss_\Theta \frac{\D p(x|\theta)}{\D p_\pi (x)} \rho_\theta \D\pi(\theta)\Big|_{u=0}\notag\\
&=\frac{\D}{\D u}\bigintss_\Theta \frac{1}{\frac{\D p_\pi (x)}{\D p(x|\theta)}} \rho_\theta \D\pi(\theta)\Big|_{u=0}\notag\\
&=\bigintss_\Theta \frac{\rho_\theta\Big( k\D\pi^{'}_m(\theta)+(1-k)\delta(\theta-\theta_0)\D\theta-\D\pi^{'}_{m+1}(\theta)\Big)}{\frac{\D p_{\pi^{'}_{m+1}}}{\D p(x|\theta)}}-\notag\\
&\bigintss_\Theta \frac{\rho_\theta \D\pi^{'}_{m+1}(\theta)}{\Bigg(\frac{\D p_{\pi^{'}_{m+1}}}{\D p(x|\theta)}\Bigg)^2}\frac{\D}{\D u}\Bigg(\frac{\D p_\pi(x)}{\D p(x|\theta)}\Bigg)\Bigg|_{u=0}.
\end{flalign}
Now, the derivative with respect to $u$ in the second term can be calculated by exchanging the order of differentiation as $p(x|\theta)$ is independent of $u$. So, we have 
\benn
\frac{\D}{\D u}\Bigg(\frac{\D p_\pi(x)}{\D p(x|\theta)}\Bigg)\Bigg|_{u=0}&=&\frac{\D}{\D p(x|\theta)}\Bigg(\frac{\D p_\pi(x)}{\D u}\Bigg)\Bigg|_{u=0}\\
&=&k\frac{\D p_{\pi^{'}_m}(x)}{\D p(x|\theta)}+(1-k) \frac{\D p(x|\theta_0)}{\D p(x|\theta)}-\frac{\D p_{\pi^{'}_{m+1}}}{\D p(x|\theta)}(x).
\eenn
Plugging in \eqref{eqA13} we obtain,
\begin{flalign*}
\frac{\D}{\D u}\hat{\rho}_B^\pi(x)\Big|_{u=0}&=\bigintss_\Theta \frac{\rho_\theta\Big( k\D\pi^{'}_m(\theta)+(1-k)\delta(\theta-\theta_0)\D\theta-\D\pi^{'}_{m+1}(\theta)\Big)}{\frac{\D p_{\pi^{'}_{m+1}}}{\D p(x|\theta)}}-\\
&\bigintss_\Theta \frac{\rho_\theta \D\pi^{'}_{m+1}(\theta)}{\Bigg(\frac{\D p_{\pi^{'}_{m+1}}}{\D p(x|\theta)}\Bigg)^2}\Bigg(k\frac{\D p_{\pi^{'}_m}(x)}{\D p(x|\theta)}+(1-k) \frac{\D p(x|\theta_0)}{\D p(x|\theta)}-\frac{\D p_{\pi^{'}_{m+1}}(x)}{\D p(x|\theta)}\Bigg).
\end{flalign*}
Now, in the limit $m\rightarrow\infty$, the coefficient of $k$ vanishes in the expression above due to weak convergence, while the last term of the first integral cancels with the last term of the second integral. So, we have
\be\label{eqA14}
\lim_{m\rightarrow\infty}\frac{\D}{\D u}\hat{\rho}_B^\pi(x)\Big|_{u=0}=\lim_{m\rightarrow\infty}(1-k)\frac{\D p(x|\theta_0)}{\D p_{\pi^{'}_{m+1}}(x)}\Big(\rho_{\theta_0}-\hat{\rho_B}^{\pi^{'}_{m+1}}(x)\Big).
\ee
Applying the limit and plugging \eqref{eqA14} in \eqref{eqA13a}, we find that the second term in \eqref{eqA12b} is
\benn
&=&\lim_{m\rightarrow\infty}\int_\mathcal{X}\D p_{\pi^{'}_{m+1}}(x)\Tr\Big(f'\big(\hat{\rho}_B^{\pi^{'}_{m+1}}(x)\big)-f'\big(\hat{\rho}(x)\big)\Big)(1-k)\frac{\D p(x|\theta_0)}{\D p_{\pi^{'}_{m+1}}(x)}\Big(\rho_{\theta_0}-\hat{\rho}_B^{\pi^{'}_{m+1}}(x)\Big)\\
&=&\lim_{m\rightarrow\infty}(1-k)\int_\mathcal{X}\D p(x|\theta_0)\Tr\Big(f'\big(\hat{\rho}_B^{\pi^{'}_{m+1}}(x)\big)-f'\big(\hat{\rho}(x)\big)\Big)\Big(\rho_{\theta_0}-\hat{\rho}_B^{\pi^{'}_{m+1}}(x)\Big).
\eenn
Finally, combining both the terms of \eqref{eqA12b} and applying the limit, we find that 
\begin{flalign*}
0\leq\lim_{m\rightarrow\infty} \frac{\D}{\D u}D^f_{\hat{\rho}}(\pi)\Bigg|_{u=0}&=\lim_{m\rightarrow\infty}\int_\mathcal{X}\Big(k\D p_{\pi^{'}_{m+1}}(x)+(1-k)\D p(x|\theta_0)-\D p_{\pi^{'}_{m+1}}(x)\Big)D_f\big(\hat{\rho}_B^{\pi^{'}_{m+1}}(x),\hat{\rho}(x)\big)+\\
&(1-k)\int_\mathcal{X}\D p(x|\theta_0)\Tr\Big(f'\big(\hat{\rho}_B^{\pi^{'}_{m+1}}(x)\big)-f'\big(\hat{\rho}(x)\big)\Big)\Big(\rho_{\theta_0}-\hat{\rho}_B^{\pi^{'}_{m+1}}(x)\Big).
\end{flalign*}
This implies that
\begin{flalign*}
\lim_{m\rightarrow\infty}(1-k)\int_\mathcal{X}\D p_{\pi^{'}_{m+1}}(x)\Big)D_f\big(\hat{\rho}_B^{\pi^{'}_{m+1}}(x),\hat{\rho}(x)\big)&\leq \lim_{m\rightarrow\infty} (1-k)\int_\mathcal{X}\D p(x|\theta_0)\Tr\Big(f'\big(\hat{\rho}_B^{\pi^{'}_{m+1}}(x)\big)-f'\big(\hat{\rho}(x)\big)\Big)\cdot\\
&\Big(\rho_{\theta_0}-\hat{\rho}_B^{\pi^{'}_{m+1}}(x)\Big)+\int_\mathcal{X}\D p(x|\theta_0)D_f\big(\hat{\rho}_B^{\pi^{'}_{m+1}}(x),\hat{\rho}(x)\big).
\end{flalign*}
The right-hand side of the inequality above can be rearranged to obtain 
\begin{flalign*}
\lim_{m\rightarrow\infty}\underbrace{\int_\mathcal{X}\D p_{\pi^{'}_{m+1}}(x)\Big)D_f\big(\hat{\rho}_B^{\pi^{'}_{m+1}}(x),\hat{\rho}(x)\big)}_\txt{$\geq0$, due to non-negativity of Bregman divergence.}&\leq \lim_{m\rightarrow\infty} \int_\mathcal{X}\D p(x|\theta_0) D_f\big(\rho_{\theta_0},\hat{\rho}(x)\big)-D_f\big(\rho_{\theta_0},\hat{\rho}_B^{\pi^{'}_{m+1}}(x)\big).
\end{flalign*}
This implies that 
\benn
\lim_{m\rightarrow\infty}R(\rho_{\theta_0},\hat{\rho}_B^{\pi^{'}_{m+1}})\leq R(\rho_{\theta_0},\hat{\rho}),~~~~~\forall\theta_0\in\Theta.
\eenn
But, as Bregman divergence is lower semi-continuous (\cref{lsc}), we have
\benn
R(\rho_{\theta_0},\lim_{m\rightarrow\infty}\hat{\rho}_B^{\pi^{'}_{m+1}})\leq \lim_{m\rightarrow\infty}R(\rho_{\theta_0},\hat{\rho}_B^{\pi^{'}_{m+1}}).
\eenn
Therefore, we arrive at our result, i.e.
\benn
R(\rho_{\theta_0},\lim_{m\rightarrow\infty}\hat{\rho}_B^{\pi^{'}_{m+1}})\leq R(\rho_{\theta_0},\hat{\rho}),~~~~~\forall\theta_0\in\Theta.
\eenn
\epr
\subsection{A Bayesian method for minimax state estimation}
Formally, \cref{thm2} is stated as the following theorem.
\begin{theorem}
There exists a convergent sequence of priors $(\pi_n)_n$ such that the limit of the sequence maximizes the average risk, Equation \eqref{<risk>}, of the Bayes estimator. The limit of such a sequence is referred to as a least favourable prior. Moreover, the sequence of Bayes estimators $(\hat{\rho}_B^{\pi_n})_n$ converges such that the limit of the sequence is minimax, i.e 
\benn
\inf_{\hat{\rho}}\sup_\theta R(\rho_\theta,\hat{\rho})=\sup_\theta R(\rho_\theta,\lim\limits_{n\rightarrow\infty}\hat{\rho}_B^{\pi_n}).
\eenn
\end{theorem}
\bpr
Consider the average risk of the Bayes estimator for a prior $\pi\in\mathcal{P}(\Theta)$ as the map
\benn 
h:\pi\mapsto r(\pi,\hat{\rho}_B^\pi)=\int_\Theta \D\pi(\theta)\int_\mathcal{X}\D p(x|\theta) D_f \big(\rho_\theta,\hat{\rho}_B^\pi(x)\big).
\eenn
Due to the discontinuity of the Bayes estimator, we follow the same regularization arguments as made earlier and define closed subsets of $\mathcal{P}(\Theta)$, Equation \eqref{subset1}, such that the Bayes estimator is continuous on each of these subsets. The map $h:\pi\mapsto r(\pi,\hat{\rho}_B^\pi)$ is then continuous on each of the subsets. Since these  subsets are closed subsets of a compact set they are themselves compact. Therefore, h attains a maximum on each of the subsets. Then, denoting the maxima in each subset $\mathcal{P}_{\mu/{n_{m+1}}}$ as $\pi^{'}_{m+1}$, we define a prior as done earlier in Equation \eqref{subset2} as a convex sum of $\pi^{'}_{m+1}$ and another element in $\mathcal{P}_{\mu/{n_{m+1}}}$. Since the average risk is maximized on $\mathcal{P}_{\mu/{n_{m+1}}}$,  the derivative of $r(\pi,\hat{\rho}_B^\pi)$ is negative as one approaches $\pi^{'}_{m+1}$, i.e.
\begin{flalign}\label{eq20}
0\geq\frac{\D}{\D u}
r(\pi,\hat{\rho}_B^\pi)\Bigg|_{u=0}&=\frac{\D}{\D u}\int_\Theta \D\pi(\theta)\int_\mathcal{X}\D p(x|\theta) D_f \big(\rho_\theta,\hat{\rho}_B^\pi(x)\big)\Bigg|_{u=0}\notag\\
&=\int_\Theta \frac{\D\pi(\theta)}{\D u}\Bigg|_{u=0}\int_\mathcal{X} \D p(x|\theta) D_f\big(\rho_\theta,\hat{\rho}_B^{\pi^{'}_{m+1}}(x)\big)+\notag\\
&\int_\Theta \D\pi^{'}_{m+1}(\theta)\int_\mathcal{X} \D p(x|\theta)\frac{\D}{\D u}D_f\big(\rho_\theta,\hat{\rho}_B^\pi(x)\big)\Bigg|_{u=0}.
\end{flalign}
Evaluating the derivatives using \cref{lemC1}, the first term in \eqref{eq20} is
\benn
\int_\mathcal{X} \D p(x|\theta_0)D_f\big(\rho_{\theta_0},\hat{\rho}_B^{\pi^{'}_{m+1}}(x)\big) - \int_\Theta  \D\pi^{'}_{m+1}(\theta)\int_\mathcal{X}\D p(x|\theta)D_f\big(\rho_\theta,\hat{\rho}_B^{\pi^{'}_{m+1}}(x)\big),
\eenn
while the derivative in the second term of \eqref{eq20} is
\begin{flalign*}
\begin{split}
\frac{\D}{\D u}D_f\big(\rho_\theta,\hat{\rho}_B^\pi(x)\big)\Bigg|_{u=0}=&\frac{\D}{\D u}\Tr\Big[f(\rho_\theta)-f(\hat{\rho}_B^\pi(x))-f'(\hat{\rho}_B^\pi(x))\big(\rho_\theta-\hat{\rho}_B^\pi(x)\big)\Big]\Bigg|_{u=0}\\
=&\Tr\Big[-f'(\hat{\rho}_B^{\pi^{'}_{m+1}}(x))\frac{\D}{\D u}\hat{\rho}_B^\pi(x)\Big|_{u=0}+f'(\hat{\rho}_B^{\pi^{'}_{m+1}}(x))\frac{\D}{\D u}\hat{\rho}_B^\pi(x)\Big|_{u=0}\\
-&\big(\rho_\theta-\hat{\rho}_B^{\pi^{'}_{m+1}}(x)\big)\frac{\D}{\D u}f'(\hat{\rho}_B^\pi(x))\Big|_{u=0}\Big]\\
=&\Tr\Big[\big(\hat{\rho}_B^{\pi^{'}_{m+1}}(x)-\rho_\theta\big)\frac{\D}{\D u}f'(\hat{\rho}_B^\pi(x))\Big|_{u=0}\Big].
\end{split}
\end{flalign*}
So, plugging this in the second term of \eqref{eq20}, we have
\begin{flalign}\label{eq18}
&\int_\Theta \D\pi^{'}_{m+1}(\theta) \int_\mathcal{X}\D p(x|\theta)\Tr\Big[\big(\hat{\rho}_B^{\pi^{'}_{m+1}}(x)-\rho_\theta\big)\frac{\D}{\D u}f'(\hat{\rho}_B^\pi(x))\Big|_{u=0}\Big]\notag\\
&=\Tr\Bigg[\int_\mathcal{X}\int_\Theta \D\pi^{'}_{m+1}(\theta)\D p(x|\theta)\big(\hat{\rho}_B^{\pi^{'}_{m+1}}(x)-\rho_\theta\big)\frac{\D}{\D u}f'(\hat{\rho}_B^\pi(x))\Big|_{u=0}\Bigg]\notag\\
&=\Tr\Bigg[\int_\mathcal{X}\Bigg(\D p_{\pi^{'}_{m+1}}(x)\hat{\rho}_B^{\pi^{'}_{m+1}}(x)-\int_\Theta \D\pi^{'}_{m+1}(\theta)\D p(x|\theta)\rho_\theta \Bigg) \frac{\D}{\D u}f'(\hat{\rho}_B^\pi(x))\Big|_{u=0}\Bigg]=0.
\end{flalign}
Thus, applying the limit $m\rightarrow\infty$ we arrive at the following inequality,
\benn 
0\geq \lim_{m\rightarrow\infty}\int_\mathcal{X} \D p(x|\theta_0)D_f\big(\rho_{\theta_0},\hat{\rho}_B^{\pi^{'}_{m+1}}(x)\big) - \int_\Theta  \D\pi^{'}_{m+1}(\theta)\int_\mathcal{X}\D p(x|\theta)D_f\big(\rho_\theta,\hat{\rho}_B^{\pi^{'}_{m+1}}(x)\big),
\eenn
which implies that
\benn
\lim_{m\rightarrow\infty}\int_\mathcal{X} \D p(x|\theta_0)D_f\big(\rho_{\theta_0},\hat{\rho}_B^{\pi^{'}_{m+1}}(x)\big)&\leq&\lim_{m\rightarrow\infty} \int_\Theta  \D\pi^{'}_{m+1}(\theta)\int_\mathcal{X}\D p(x|\theta)D_f\big(\rho_\theta,\hat{\rho}_B^{\pi^{'}_{m+1}}(x)\big).
\eenn
So, we have
\benn
\lim_{m\rightarrow\infty}R(\rho_{\theta_0},\hat{\rho}_B^{\pi^{'}_{m+1}})&\leq&\lim_{m\rightarrow\infty}\int_\Theta  \D\pi^{'}_{m+1}(\theta)R(\rho_\theta,\hat{\rho}_B^{\pi^{'}_{m+1}}),~~~\forall \theta_0\in\Theta.
\eenn
The lower semi-continuity of Bregman divergence implies that
\benn
R(\rho_{\theta_0}, \lim_{m\rightarrow\infty}\hat{\rho}_B^{\pi^{'}_{m+1}})&\leq&\lim_{m\rightarrow\infty}\int_\Theta  \D\pi^{'}_{m+1}(\theta)R(\rho_\theta,\hat{\rho}_B^{\pi^{'}_{m+1}}).
\eenn
Thus, we have 
\benn
\sup_{\theta_0}R(\rho_{\theta_0}, \lim_{m\rightarrow\infty}\hat{\rho}_B^{\pi^{'}_{m+1}})\leq\lim_{m\rightarrow\infty}\int_\Theta  \D\pi^{'}_{m+1}(\theta)R(\rho_\theta,\hat{\rho}_B^{\pi^{'}_{m+1}}).
\eenn
But, the other direction of the inequality above is true trivially i.e. 
\benn
\sup_{\theta_0}R(\rho_{\theta_0}, \lim_{m\rightarrow\infty}\hat{\rho}_B^{\pi^{'}_{m+1}})\geq\lim_{m\rightarrow\infty}\int_\Theta  \D\pi^{'}_{m+1}(\theta)R(\rho_\theta,\hat{\rho}_B^{\pi^{'}_{m+1}}).
\eenn
Therefore, we obtain
\be\label{eq21}
\sup_{\theta_0}R(\rho_{\theta_0}, \lim_{m\rightarrow\infty}\hat{\rho}_B^{\pi^{'}_{m+1}})=\lim_{m\rightarrow\infty}\int_\Theta  \D\pi^{'}_{m+1}(\theta)R(\rho_\theta,\hat{\rho}_B^{\pi^{'}_{m+1}}).
\ee
But, 
\benn
\lim_{m\rightarrow\infty}\int_\Theta  \D\pi^{'}_{m+1}(\theta)R(\rho_\theta,\hat{\rho}_B^{\pi^{'}_{m+1}})&=&\lim_{m\rightarrow\infty} \sup_{\pi\in\mathcal{P}_{\mu/{n_{m+1}}}}\int_\Theta \D\pi(\theta) R(\rho_\theta,\hat{\rho}_B^\pi).
\eenn
By \cref{lemE2}, the limit of the suprema over subsets $\mathcal{P}_{\mu/{n_{m+1}}}$ can be replaced by a supremum over the set $\mathcal{P}(\Theta)$ since the sequence of subsets $\mathcal{P}_{\mu/{n_{m+1}}}$ is dense in $\mathcal{P}(\Theta)$. Thus, we have
\be\label{eq22}
\lim_{m\rightarrow\infty}\int_\Theta  \D\pi^{'}_{m+1}(\theta)R(\rho_\theta,\hat{\rho}_B^{\pi^{'}_{m+1}})&=&\sup_{\pi\in\mathcal{P}(\Theta)}\int_\Theta \D\pi(\theta) R(\rho_\theta,\hat{\rho}_B^\pi)\notag\\
&=&\sup_{\pi\in\mathcal{P}(\Theta)}\inf_{\hat{\rho}}\int_\Theta \D\pi(\theta) R(\rho_\theta,\hat{\rho}).
\ee
Using the minimax theorem for lower semi-continuous and quasi-convex functions~\cite[Theorem 3.4]{sion1958}, we can exchange the infimum and the supremum to obtain
\benn
\sup_{\pi\in\mathcal{P}(\Theta)}\inf_{\hat{\rho}}\int_\Theta \D\pi(\theta) R(\rho_\theta,\hat{\rho})=\inf_{\hat{\rho}}\sup_{\pi\in\mathcal{P}(\Theta)}\int_\Theta \D\pi(\theta) R(\rho_\theta,\hat{\rho})=\inf_{\hat{\rho}}\sup_{\Theta}R(\rho_\theta,\hat{\rho}).
\eenn
Thus, by \eqref{eq21} and \eqref{eq22} we have the result
\benn
\inf_{\hat{\rho}}\sup_\theta R(\rho_\theta, \hat{\rho})=\sup_{\theta}R(\rho_\theta, \lim_{m\rightarrow\infty}\hat{\rho}_B^{\pi^{'}_{m+1}}).
\eenn
\epr
\cref{thm1} and \cref{thm2} are based on the assumption that the underlying POVM is fixed. However, in general, the risk depends on the POVM P, i.e. $R(\rho_\theta,\hat{\rho})\equiv R_P(\rho_\theta,\hat{\rho})$.  One way of defining an optimal POVM could be to minimize the worst-case risk over $\mathscr{P}$, the convex set of all POVMs. The POVM that minimizes the worst-case risk is called a \textit{minimax} POVM.
\begin{definition}[Minimax POVM]\label{def4}
A POVM $P^*$ is minimax if:
\be\label{eq33}
\inf_{P\in \mathscr{P}}\inf_{\hat{\rho}}\sup_\theta R_P(\rho_\theta,\hat{\rho})=\inf_{\hat{\rho}}\sup_\theta R_{P^*}(\rho_\theta,\hat{\rho})
\ee
where  $\rho_\theta$ is the estimand, $\hat{\rho}:\mathcal{X}\mapsto\mathcal{S}(\mathcal{H})$ is an estimator with risk $R_P(\rho_\theta,\hat{\rho})$ which is a function of the POVM P, and $\mathscr{P}$ is the convex set of all POVMs on the measurement outcome space $\mathcal{X}$.
\end{definition}
It remains unclear as to how one could obtain a minimax POVM for general state estimation, but if we restrict ourselves to the case of \emph{covariant state estimation} the situation simplifies.
\subsection{Covariant state estimation}
In the covariant state estimation problem, as discussed in reference \cite{holevo82}, one is given a fixed state $\rho_{\theta_0}$ and is interested in estimating all the states $\rho_\theta$ that lie in the orbit $\{V_g\rho_{\theta_0}V_g^\dagger\}$, where $g\in G$ is a parametric group of transformations of the parameter space $\Theta$ and $g\mapsto V_g$ is a (continuous) projective unitary representation of G. One can think of this as representing the following physical scenario. Given that the parameter $\theta$ labels the quantum states of the Hilbert space $\mathcal{H}$, $\theta$ can be assumed to be describing some aspects of the preparation procedure for the state $\rho_\theta$---a transformation $g$ of the parameter $\theta_0$ results in the preparation of the state $\rho_\theta=V_g\rho_{\theta_0}V_g^\dagger$ where $\theta=g\theta_0$. Covariant state estimation thus corresponds to the estimation of the state $\rho_\theta$ with the measurement outcome space $\mathcal{X}$ being identical to the parameter space $\Theta$. Let us first define a \emph{covariant} measurement.
\begin{definition}[Covariant measurement]\label{def5}
Let G be a parametric group of transformations of a set $\Theta$ and $g\mapsto V_g$ be a (continuous) projective unitary representation of G in a Hilbert space $\mathcal{H}$. Let $M(\D\hat{\theta})$ be a \textit{positive operator-valued measure} defined on the $\sigma$-algebra $\mathcal{A}(\Theta)$ of Borel subsets of $\Theta$. Then $M(\D\hat{\theta})$ is \textit{covariant} with respect to the representation $g\mapsto V_g$ if 
\be\label{M_c}
V_g^\dagger M(B) V_g=M(B_{g^{-1}}), ~~~~~g\in G,
\ee
for any $B\in\mathcal{A}(\Theta)$, where $B_g=\{\theta': \theta'=g\theta,~~\theta\in B\}$.
\end{definition}
If $M(d\hat{\theta})$ is covariant, then
\benn
\Tr M(B)\rho_\theta=\Tr\rho_0 V_g^\dagger M(B)V_g=\Tr\rho_0 M(B_{g^{-1}}).
\eenn
Thus, 
\benn
\txt{Pr}[\hat{\theta}\in B|~g\theta_0]=\txt{Pr}[\hat{\theta}\in B_{g^{-1}}|~\theta_0],
\eenn
i.e. a covariant measurement preserves the probability distribution under the transformation of the state. We refer the reader to reference \cite{holevo82} for a more detailed discussion on covariant measurements.

Before we start building towards the proof of \cref{thm3}, let us look at some of the properties of the parametric group G to understand the situation better. First, the group G is chosen to act transitively on $\Theta$. This ensures that the map $g\mapsto g\theta_0$ maps G onto the whole $\Theta$. Second, G is assumed to be unimodular which implies that there exists an invariant measure $\mu$ on G. Third, G is assumed to be compact which ensures that the measure $\mu<\infty$. The measure $\mu$ is normalized as $\mu(G)=1$. Now, we are interested in an invariant measure $\nu$ on the $\Sigma$-algebra $\mathcal{A}(\Theta)$ on $\Theta$ such that \[\nu(B)=\nu(B_g),~~~B\in\mathcal{A}(\Theta),\] where $B_g=\{\theta': \theta'=g\theta,~~\theta\in B\}$. If $G_0$, the stationary subgroup of G, is unimodular then such a measure $\nu$ exists and if $G_0$ is compact then $\nu$ is finite and can be constructed from $\mu$ by demanding that the following relation holds for all integrable functions f on $\Theta$:
\be
\int_G f(g\theta_0)\D\mu(g)=\int_\Theta f(\theta)\D\nu(\theta).
\ee
We now state Proposition 2.1 from reference \cite{holevo82} as the following lemma that gives a relation between the two measures.
\begin{lemma}\label{lem3}
Let $M(d\theta)$ be a measurement covariant with respect to a projective unitary representation $g\mapsto V_g$ of the parametric group G acting on $\Theta$. For any density operator $\rho \in \mathcal{H}$, and for any Borel set $B\in\mathcal{A}(\Theta)$
\be\label{45}
\int_G \Tr[V_g\rho V_g^\dagger M(B)]\D\mu(g)=\nu(B)
\ee
\end{lemma}
Let us pause here to look at an example.
\begin{eg}\label{eg3}
Let us assume that we are interested in estimating all those states in $\mathbb{C}^2$ that lie on the Bloch sphere. Thus, the parameter space $\Theta$ is $\mathbb{S}^2$. This is a covariant estimation problem with the parametric group of transformations on $\mathbb{S}^2$ being SO(3). Its projective unitary representation is the quotient subgroup SU(2)/U(1). Let us assume that the initial state $\rho_{\theta_0}$ is $|0\X0|$. Then, the elements of SU(2)/U(1) generate all the states on the Bloch sphere that are parametrized by a set of two parameters: the latitude $\theta$\footnote{We apologize for the redundancy, but it is best to stick to conventional symbols.} and azimuth $\phi$, with the states being identified as $|\theta,\phi\rangle$. Note that an element of SU(2)/U(1) can be written in terms of these as
\benn
U_{\theta,\phi}=\begin{bmatrix}
				\cos{\frac{\theta}{2}} & -\sin{\frac{\theta}{2}}e^{-i\phi}\\
				\sin{\frac{\theta}{2}}e^{i\phi} & \cos{\frac{\theta}{2}}
				\end{bmatrix}.
\eenn
It can be shown \cite[Theorem 2.1]{holevo82} (stated as \cref{lemE3} for reference) that starting from a positive operator $P_0$ that commutes with all the elements of the stationary subgroup of SU(2)/U(1) such that it satisfies Equation \eqref{lemE328}, setting $M(\theta,\phi)=U_{\theta,\phi}P_0U_{\theta,\phi}^\dagger$ implies that the measurement defined with $M(\theta,\phi)$ as the operator-valued density with respect to the uniform measure on $\Theta$ is covariant. In this case, $P_0=2|0\X0|$ and the operator-valued density is 
\benn
M(\theta,\phi)=2\begin{bmatrix}
				\cos{\frac{\theta}{2}}\\
				\sin{\frac{\theta}{2}}e^{i\phi}
			  \end{bmatrix}\cdot\begin{bmatrix}
							\cos{\frac{\theta}{2}} & \sin{\frac{\theta}{2}}e^{-i\phi}
						   \end{bmatrix}=2\begin{bmatrix}
											\cos^2{\frac{\theta}{2}} & \cos{\frac{\theta}{2}}\sin{\frac{\theta}{2}}e^{-i\phi}\\
											\cos{\frac{\theta}{2}}\sin{\frac{\theta}{2}}e^{i\phi} & \sin^2{\frac{\theta}{2}}
										   \end{bmatrix}.
\eenn 
It is straightforward to verify that $\frac{1}{4\pi}\int_{\mathbb{S}^2}M(\theta,\phi)\sin{\theta}\D\theta\D\phi=\mathbb{I}$. 
\end{eg}
Having defined and illustrated the problem of covariant state estimation, we now recall Holevo's theorem \cite[Theorem 3.1]{holevo82}, which states that for every loss function that is invariant under the group transformation $g$, the minimax risk as well as the average risk attain their minima at a covariant measurement. But, the analysis in reference \cite{holevo82} is done for loss functions expressed as functions of the true parameter and the estimator of the parameter. It can be recast in terms of the general framework involving estimators that are functions of the parameter, $\hat{\rho}:\Theta\mapsto\mathcal{D}(\mathcal{H})$ by simply choosing the domain of the loss function to be the set of density matrices $\mathcal{S}(\mathcal{H})$ as opposed to the parameter space $\Theta$. We thus state it as the following lemma which is the main ingredient of the proof of \cref{thm3}.


\begin{lemma}[Theorem 3.1, \cite{holevo82}]\label{Holevo}
In the quantum covariant statistical estimation problem, given an estimator $\hat{\rho}$ of the state, the minima of the average risk $r_P(\nu,\hat{\rho})$ with respect to the uniform Haar measure $\nu$ on $\Theta$ and the worst case risk $\sup_\theta R_P(\rho_\theta,\hat{\rho})$ for all $\Theta-$measurements are achieved on a covariant measurement. Moreover, for any covariant measurement $P_c$, we have
\be
r_{P_c}(\nu,\hat{\rho})=\sup_\theta R_{P_c}(\rho_\theta,\hat{\rho})=R_{P_c}(\rho_\theta,\hat{\rho}),~~~~\theta\in\Theta
\ee
\end{lemma}

\emph{Note}: A covariant measurement is \emph{not} a unique minimum for either the average or the worst-case risk.

The above theorem implies that for any measurement $P$, there exists a covariant measurement that minimizes the average risk as well as the worst case risk for a fixed estimator $\hat{\rho}$. But, we know that for a fixed measurement the average risk is minimized by the Bayes estimator $\hat{\rho}_B$. Thus, the Bayes estimator minimizes the average risk for a covariant measurement. However, the invariance of the loss function implies that the Bayes estimator must be covariant under the group transformations as shown below.
\begin{lemma}
The Bayes estimator is covariant under the group transformations $V_g$, i.e. 
\benn
\hat{\rho}_B(B_g)=V_g\hat{\rho}_B(B)V_g^\dagger,~~~B\in\mathcal{A}(\Theta).
\eenn
\end{lemma}
\bpr
Recalling the invariance property of the loss function: $L(\rho_\theta,\hat{\rho}(x))=L(\rho_{g\theta},\hat{\rho}(gx))=L(V_g\rho_\theta V_g^\dagger,V_g\hat{\rho}(x)V_g^\dagger)$, let us verify for the case of Bayes estimator. Recalling Equation \eqref{eq13}, we have
\benn
\hat{\rho}_B(B_g)&=&\frac{\int_\Theta \D\nu(\theta)\Tr\rho_\theta P(B_g)\rho_\theta}{\int_\Theta \D\nu(\theta)\Tr\rho_\theta P(B_g)}.
\eenn
As we are interested in a covariant measurement $P(B_g)=V_gP(B)V_g^\dagger$, therefore
\benn
\hat{\rho}_B(B_g)&=&\frac{\int_\Theta \D\nu(\theta)\rho_\theta\Tr\rho_\theta V_gP(B)V_g^\dagger}{\int_\Theta \D\nu(\theta)\Tr\rho_\theta V_gP(B)V_g^\dagger}\\
&=&\frac{\int_\Theta \D\nu(\theta)\rho_\theta\Tr V_g^\dagger\rho_\theta V_gP(B)}{\int_\Theta \D\nu(\theta)\Tr V_g^\dagger\rho_\theta V_gP(B)}\\
&=&\frac{\int_\Theta \D\nu(\theta)\rho_\theta\Tr \rho_{g^{-1}\theta} P(B)}{\int_\Theta \D\nu(\theta)\Tr \rho_{g^{-1}\theta}P(B)}.
\eenn
By the invariance of $\nu$ and the fact that $\rho_\theta=V_g\rho_{g^{-1}\theta}V_g^\dagger$, we finally obtain
\benn
\hat{\rho}_B(B_g)&=&\frac{\int_\Theta \D\nu(g^{-1}\theta)V_g\rho_{g^{-1}\theta}V_g^\dagger\Tr \rho_{g^{-1}\theta} P(B)}{\int_\Theta \D\nu(g^{-1}\theta)\Tr \rho_{g^{-1}\theta}P(B)}\\
&=&V_g\Bigg\{\frac{\int_\Theta \D\nu(g^{-1}\theta)\rho_{g^{-1}\theta}\Tr \rho_{g^{-1}\theta} P(B)}{\int_\Theta \D\nu(g^{-1}\theta)\Tr \rho_{g^{-1}\theta}P(B)}\Bigg\}V_g^\dagger=V_g\hat{\rho}_B(B)V_g^\dagger.
\eenn
\epr
Thus, we have established two things about a covariant measurement. First, that the risk of a covariant measurement is independent of the state $\rho_\theta$ and second, that it minimizes the average as well as the worst-case risk among all measurements. But, the problem of finding a minimax POVM is closely tied to obtaining a least favourable prior which in turn is tied to the underlying measurement, as we discussed in \cref{sec2}. The following lemma gives a least favourable prior and the corresponding measurement in the context of covariant state estimation.
\Covlemma*
\bpr
As the Bayes estimator $\hat{\rho}_B^{\pi}$ minimizes the average risk with respect to the prior $\pi$, 
\benn
\sup_\theta R_{P_c}(\rho_\theta, \rho_B^{\nu})=\sup_\pi\int_\Theta \D\pi(\theta)R_{P_c}(\rho_\theta, \rho_B^{\nu})\geq\sup_\pi\int_\Theta\txt{d}\pi(\theta)R_{P_c}(\rho_\theta, \rho_B^{\pi}).
\eenn
However, by \cref{Holevo}, we know that for a covariant measurement the risk is independent of the state $\rho_\theta$, i.e.
\benn
\sup_\theta R_{P_c}(\rho_\theta, \rho_B^{\nu})&=&\int_\Theta \txt{d}\nu(\theta)R_{P_c}(\rho_\theta, \rho_B^{\nu}).
\eenn
This implies that
\benn
\int_\Theta \txt{d}\nu(\theta)R_{P_c}(\rho_\theta, \rho_B^{\nu})&\geq&\sup_\pi\int_\Theta\txt{d}\pi(\theta)R_{P_c}(\rho_\theta, \rho_B^{\pi}).
\eenn
The other direction of the above inequality holds trivially, i.e.
\benn
\int_\Theta \txt{d}\nu(\theta)R_{P_c}(\rho_\theta, \rho_B^{\nu})&\leq&\sup_\pi\int_\Theta\txt{d}\pi(\theta)R_{P_c}(\rho_\theta, \rho_B^{\pi}).
\eenn
Therefore,
\benn
\int_\Theta \txt{d}\nu(\theta)R_{P_c}(\rho_\theta, \rho_B^{\nu})=\sup_\pi\int_\Theta\txt{d}\pi(\theta)R_{P_c}(\rho_\theta, \rho_B^{\pi}),
\eenn
and so $\nu$ is a least favourable prior for a covariant measurement $P_c$.
\epr
The only remaining ingredient needed to prove \cref{thm3} is the following lemma.
\begin{lemma}\label{lem6}
The Bayes estimator for a covariant measurement $P_c$ is 
\be
\hat{\rho}_B(B)=\frac{1}{\nu(B)}\Tr\Big[\mathbb{I}^R\otimes P_c^{R'}(B)\int_G \D\mu(g)\big(V_g\rho_0 V_g^\dagger\big)^{\otimes 2}\Big], ~~~B\in\mathcal{A}(\Theta).
\ee
\end{lemma}
\bpr
 Recalling Equation \eqref{eq13}, the Bayes estimator for a covariant measurement $P_c$ is 
\benn
\hat{\rho}_B(B)&=&\frac{\int_\Theta \D\nu(\theta)\Tr[\rho_\theta P_c(B)]\rho_\theta}{\int_\Theta \D\nu(\theta)\Tr\rho_\theta P_c(B)}, ~~~B\in\mathcal{A}(\Theta).
\eenn
Using \cref{lem3}, the denominator in the above expression is
\benn
\int_\Theta \D\nu(\theta)\Tr\rho_\theta P_c(B)=\nu(B),
\eenn
while the numerator is
\benn
\int_\Theta \D\nu(\theta)\Tr[\rho_\theta P_c(B)]\rho_\theta&=&\int_G \D\mu(g)\Tr[ \rho_{g\theta_0}P_c(B)]\rho_{g\theta_0}\\
&=&\int_G \D\mu(g) \Tr[V_g\rho_0 V_g^\dagger P_c(B)]V_g \rho_0 V_g^\dagger\\
&=&\Tr_{R'}\Big[\mathbb{I}^R\otimes P_c^{R'}(B)\int_G \D\mu(g)\big(V_g\rho_0 V_g^\dagger\big)^R\otimes \big(V_g\rho_0 V_g^\dagger\big)^{R'}\Big].
\eenn
\epr
Now that we have a class of measurements and the corresponding least favourable prior, in order to show that it is minimax (first part of \cref{thm3}), we have to show that this class of measurements also minimizes the average risk with respect to such a least favourable prior. Recall \cref{thm3}.
Formally, the first part of \cref{thm3} is stated as the following theorem and the second part as a corollary to the theorem.
\begin{theorem}\label{thm3*}
There exists a covariant measurement that is minimax for covariant state estimation.
\end{theorem}
\bpr
Recalling \cref{def4} of a minimax POVM and the fact that the Bayes estimator minimizes the average risk, we have
\benn
\inf_P \inf_{\hat{\rho}}\sup_\theta R_P(\rho_\theta,\hat{\rho})&=&\inf_P \inf_{\hat{\rho}}\sup_\pi\int_\Theta \txt{d}\pi(\theta)R_P(\rho_\theta,\hat{\rho})\\
&\geq& \inf_P \sup_\pi\int_\Theta \txt{d}\pi(\theta)R_P(\rho_\theta,\rho_B^\pi).
\eenn
But, as \[\sup_\pi\int_\Theta \txt{d}\pi(\theta)R_P(\rho_\theta,\rho_B^\pi)\geq \int_\Theta \txt{d}\mu(\theta)R_P(\rho_\theta,\rho_B^{\mu}),~~\forall\mu\in\mathcal{P}(\Theta),\] it implies that the same holds for the uniform Haar measure $\nu$ as well. Thus,
\benn
\inf_P \inf_{\hat{\rho}}\sup_\theta R_P(\rho_\theta,\hat{\rho})\geq\inf_P \int_\Theta \txt{d}\nu(\theta)R_P(\rho_\theta,\rho_B^{\nu}).
\eenn
Now, we know from \cref{Holevo} that 
\benn
\inf_P \int_\Theta \txt{d}\nu(\theta)R_P(\rho_\theta,\rho_B^{\nu})=\int_\Theta \txt{d}\nu(\theta)R_{P_c}(\rho_\theta,\rho_B^{\nu}),
\eenn
where $P_c$ is a covariant measurement that minimizes the average risk. Also, by \cref{lem2}, $\nu$ is a least favourable prior which means that $\hat{\rho}_B^{\nu}$ is a minimax estimator. Therefore, we have
\benn
\int_\Theta \txt{d}\nu(\theta)R_{P_c}(\rho_\theta, \rho_B^{\nu})=\sup_\theta R_{_c}(\rho_\theta, \rho_B^{\nu})=\inf_{\hat{\rho}}\sup_\theta R_{P_c}(\rho_\theta, \hat{\rho}).
\eenn
Thus, we obtain
\benn
\inf_P \inf_{\hat{\rho}}\sup_\theta R_P(\rho_\theta,\hat{\rho})\geq\inf_{\hat{\rho}}\sup_\theta R_{P_c}(\rho_\theta, \hat{\rho}).
\eenn
The other direction of the inequality above holds trivially, i.e.
\benn
\inf_P \inf_{\hat{\rho}}\sup_\theta R_P(\rho_\theta,\hat{\rho})\leq\inf_{\hat{\rho}}\sup_\theta R_{P_c}(\rho_\theta, \hat{\rho})
\eenn
Hence, we have proved that $P_c$ is a minimax POVM, i.e. 
\benn
\inf_P \inf_{\hat{\rho}}\sup_\theta R_P(\rho_\theta,\hat{\rho})=\inf_{\hat{\rho}}\sup_\theta R_{P_c}(\rho_\theta, \hat{\rho}).
\eenn
\epr
As the risk for a covariant measurement is independent of the state $\rho_\theta$ (by \cref{Holevo}) and depends only on the estimator (the Bayes estimator in this case, which is a function of $\int_G \D\mu(g)\big(V_g\rho_0 V_g^\dagger\big)^{\otimes 2}$), we have the following corollary.
\begin{corollary}
Given a minimax covariant measurement $\mathscr{P}_c$, if there exists a measurement $\mathscr{P}^\prime_c$ which is covariant under a subgroup H of G such that $\{V_h|~h\in H\}$, where $V_h$ is the projective unitary representation of the subgroup H, forms a unitary 2-design, i.e. 
\benn
\mathscr{P}_c(B)=V_h^\dagger \mathscr{P}_c(B_{g^{-1}})V_h,~~~B\in\mathcal{A}(\Theta);~h\in H,
\eenn
where $B_{g^{-1}}=\{g^{-1}\theta|~\theta\in B\}$, and $\mathscr{P}_c$ and $\mathscr{P}^\prime_c$ have the same seed, then $\mathscr{P}^\prime_c$ is also minimax.
\end{corollary}
In order to understand the above corollary better let us look at what it means for \cref{eg3}.
\begin{eg3contd}\label{eg3contd}
Now, since any state $|\theta,\phi\rangle$ in $\mathbb{S}^2$ can be generated by elements of SU(2)/U(1), the following equivalence holds:
\benn
\frac{1}{4\pi}\int_{\mathbb{S}^2}|\theta,\phi\X\theta,\phi|\sin{\theta}\D\theta\D\phi=\int_{SU(2)/U(1)}U_{\theta,\phi}|0\X0|U_{\theta,\phi}^\dagger\D U_{\theta,\phi},
\eenn
where $\D U_{\theta,\phi}$ is the Haar measure on SU(2)/U(1). Infact, the above implies that
\benn
\frac{1}{4\pi}\int_{\mathbb{S}^2}\big(|\theta,\phi\X\theta,\phi|\big)^{\otimes 2}\sin{\theta}\D\theta\D\phi=\int_{SU(2)/U(1)}\big(U_{\theta,\phi}|0\X0|U_{\theta,\phi}^{\dagger}\big)^{\otimes 2}\D U_{\theta,\phi}.
\eenn
Now, the above equivalence along with  \cite[Theorem 3.3.1]{DerakhshaniQt-design} implies that one can construct a unitary 2-design from a quantum 2-design. Thus, the set of unitary matrices that generate the set of eigenstates of the Pauli matrices $\sigma_x,\sigma_y,\sigma_z$ is a unitary 2-design given as below.
\benn
U_{0,0}&=&\begin{bmatrix}
1 & 0\\
0 & 1
\end{bmatrix}, U_{\pi,0}=\begin{bmatrix}
				0 & 1\\
				1 & 0
			   \end{bmatrix}, \\ U_{\pi/2,0}&=&\frac{1}{\sqrt{2}}\begin{bmatrix}
											1 & -1\\
											1 & 1
										   \end{bmatrix}, U_{\pi/2,\pi}=\frac{1}{\sqrt{2}}\begin{bmatrix}
																		1 & 1\\
																		-1 & 1
										    							   \end{bmatrix}, \\
U_{\pi/2,\pi/2}&=&\frac{1}{\sqrt{2}}\begin{bmatrix}
					1 & i\\
					i & 1
				\end{bmatrix}, U_{\pi/2,3\pi/2}= \frac{1}{\sqrt{2}}\begin{bmatrix}
												1 & -i\\
												 -i & 1																		     \end{bmatrix}.
\eenn
The corresponding measurement that is covariant under the above unitary 2-design is then obtained via $M_{\theta,\phi}=U_{\theta,\phi}P_0U_{\theta,\phi}^\dagger$. It is straightforward to see that the corresponding $M_{\theta,\phi}$ are the same as the Pauli measurements apart from a normalization constant. This measurement is thus minimax.
\end{eg3contd}
\emph{Note:} In the above example we obtained a minimax POVM---a covariant measurement for the parameter space $\mathbb{S}^2$ which describes only pure qubit states. There is no a-priori reason to believe that the same measurement would also be minimax for estimating an arbitrary state of a qubit. However, curiously, it happens to be true for a qubit as we will see in the following section.


\section{Example: Minimax measurement for a qubit}\label{sec4}
We look at the single qubit case as studied in reference \cite{Komaki2017} wherein the authors obtain such a minimax POVM with relative entropy as the distance-measure. We generalize their results to squared-distance $\|\rho-\sigma\|^2=\Tr(\rho-\sigma)^2$. However, the proof does not follow the generalized treatment in terms of Bregman divergence as done in the previous sections.

To begin with, let us write the most general expression for a POVM on $\mathbb{C}^2$. Recalling \cref{lem0}, we can write any POVM, see \cref{def1}, as an operator-valued density, i.e.
\benn
P(B)=\int_B \mu(dx) M(x),
\eenn
where $B\in\sigma(\mathcal{X})$, $M(x)\geq0$, Tr$[M(x)]=2$ and $\mu(\mathcal{X})=1$. Since positive operators can be expanded in the Pauli basis $\{\mathbb{I},\sigma_x,\sigma_y,\sigma_z\}$ with real coefficients, $M(\vec{x})=\alpha_0\mathbb{I}+\vec{x}\cdot\vec{\sigma}$, but the trace 1 condition on $M(\vec{x})$ implies $\alpha_0=1/2$. Without loss of generality,
\benn
M(\vec{x})=(\mathbb{I}+\vec{x}\cdot\vec{\sigma}).
\eenn
Thus, the most general form of a POVM element on $\mathbb{C}^2$ is
\be\label{eq34}
P(B)&=&\int_B(\mathbb{I}+\vec{x}\cdot\vec{\sigma}) \D \mu(\vec{x}).
\ee
Now that we have obtained the general expression for a POVM on $\mathbb{C}^2$, we next evaluate the Bayes estimator which in turn is needed to evaluate the risk $R_P(\rho_\theta,\rho_B^\pi)$. Recalling that the Bayes estimator is given as
\benn
\rho_B^\pi(x)&=&\int_\Theta \frac{\D p(x|\theta)}{\D p_{\pi}(x)}\rho_\theta\D\pi(\theta).
\eenn
Recalling the differential form of Born's rule \eqref{Born'sruleDiff}, and using the Bloch sphere notation of $\rho_\theta$, $\rho_\theta=\frac{1}{2}(\mathbb{I}+\vec{\theta}\cdot\vec{\sigma})$, it is a straightforward calculation to obtain
\be\label{eq36}
\Tr M(\vec{x})\rho_\theta=\D\mu(\vec{x})(1+\vec{x}\cdot\vec{\theta}).
\ee
However, to be able to further simplify the Bayes estimator, we need to impose some restrictions on the prior $\pi$ defined on the parameter space $\Theta$ (which in this case is $\mathbb{R}^3$). In particular, we choose a uniform prior $\pi^*$ supported only on pure states, i.e. $\pi(\theta)$ is zero for all vectors $\vec{\theta}$ with $\|\theta\|<1$ but is uniformly distributed on the set of unit vectors with $\|\theta\|=1$. It can be verified that such a prior has the following two properties~:
\begin{enumerate}
\item $\mathbb{E}_{\pi^*}[\theta_i]=0,~~~\forall i\in\{x,y,z\}$.
\item $\mathbb{E}_{\pi^*}[\theta_i\theta_j]=\frac{1}{3}\delta_{ij}~~~\forall i,j\in\{x,y,z\}$.
\end{enumerate}
By property (1) of the prior $\pi^*$, we have
\be\label{eq37}
p_\pi(B)=\int_\Theta\int_B\txt{d}\mu(\vec{x})(1+\vec{x}\cdot\vec{\theta})\pi^*(\theta)\txt{d}\theta=\mu(B). 
\ee
Thus, the Bayes estimator reduces to
\benn
\rho_B^{\pi^*}(\vec{x})&=&\bigintsss_\Theta \frac{1}{\frac{\D p_{\pi^*}(x)}{\D p(x|\theta)}}\rho_\theta\D\pi^*(\theta)\\
&=&\bigintsss_\Theta \frac{1}{\frac{\D \mu(x)}{\D p(x|\theta)}}\rho_\theta\D\pi^*(\theta)\\
&=& \bigintsss_\Theta \frac{\D p(x|\theta)}{\D\mu(x)}\rho_\theta \D\pi^*(\theta)\\
&=& \int_\Theta \rho_\theta\D\pi^*(\theta) \Tr M(x)\rho_\theta\\
&=& \frac12\int_\Theta \D\pi^*(\theta) (1+\vec{x}\cdot\vec{\theta}) (\mathbb{I}+\vec{\theta}\cdot\vec{\sigma})\\
&=& \frac12\Big(\mathbb{I}+\int_\Theta \D\pi^*(\theta)(\vec{x}\cdot\vec{\theta})(\vec{\theta}\cdot\vec{\sigma}))\Big)\\
&=& \frac12\Big(\mathbb{I}+\frac13\vec{x}\cdot\vec{\sigma}\Big).
\eenn

Now, recalling that the risk is a function of the POVM P, i.e.
\be\label{risk}
R_P(\rho_\theta, \rho_B^{\pi^*})=\int_{\mathcal{X}}\D\mu(\vec{x})\Tr M(\vec{x})\rho_\theta~D_f(\rho_\theta,\rho_B^{\pi^*}(\vec{x})),
\ee
we evaluate the risk for both relative entropy and Hilbert-Schmidt distance below.

\begin{lemma}\label{risklem}
The risk for relative entropy and Hilbert-Schmidt distance are given as
\begin{flalign*}
R_P^{rel}(\rho_\theta, \rho_B^{\pi^*})&=-h\Bigg(\frac{1+\|\theta\|}{2}\Bigg)+\frac{1}{2}\log\frac{9}{2}-\frac{\log2}{2}\int_{\mathcal{X}}\D\mu(\vec{x})\sum\limits_{j,k}\theta_j\theta_k x_j x_k, ~~~\txt{and}\\
R_P^{sq}(\rho_\theta, \rho_B^{\pi^*})&=\frac{1}{2}\Big\{\|\vec{\theta}\|^2+\frac{1}{9}\int_{\mathcal{X}}\D\mu(\vec{x})\|\vec{x}\|^2+\frac{1}{9}\int_{\mathcal{X}}\D\mu(\vec{x})\|\vec{x}\|^2\vec{x}\cdot\vec{\theta}-\frac{2}{3}\int_{\mathcal{X}}\D\mu(\vec{x})\sum_{j,k}\theta_j\theta_kx_j x_k\Big\},
\end{flalign*}
respectively.
\end{lemma}
\bpr
(a) For \textit{relative entropy}, see \cite[pg.~11]{Komaki2017}.\\
(b) For~\textit{Hilbert-Schmidt distance}, substituting $f=\|A\|_1^2$ in \eqref{risk}, we get
\benn
D_{sq}(\rho_\theta,\rho_B^{\pi^*}(x))&=&\Tr\big(\rho_\theta-\rho_B^{\pi^*}(x)\big)^2.
\eenn
Thus, 
\benn
D_{sq}(\rho_\theta,\rho_B^{\pi^*}(B))&=&\frac{1}{2}\Tr\Bigg[
\begin{bmatrix}
1+\theta_z & \theta_x-i\theta_y\\
\theta_x+i\theta_y & 1-\theta_z
\end{bmatrix}-
\begin{bmatrix}
1+z/3 & (x-iy)/3\\
(x+iy)/3 & 1-z/3
\end{bmatrix}\Bigg]^2\\
&=&\frac{1}{2}\Tr\Big[(\theta_z-z/3)\sigma_z+(\theta_x-x/3)\sigma_x+(\theta_y-y/3)\sigma_y\Big]^2\\
&=&\frac{1}{2}\Big((\theta_z-z/3)^2+(\theta_x-x/3)^2+(\theta_y-y/3)^2\Big)\\
&=&\frac{1}{2}\Big(\|\vec{\theta}\|^2+\frac{1}{9}\|\vec{x}\|^2-\frac{2}{3}\vec{x}\cdot\vec{\theta}\Big).
\eenn
This implies that the risk is
\benn
R_P^{sq}(\rho_\theta, \rho_B^{\pi^*})=\frac{1}{2}\int_{\mathcal{X}}\D\mu(\vec{x})(1+\vec{x}\cdot\vec{\theta})\Big(\|\vec{\theta}\|^2+\frac{1}{9}\|\vec{x}\|^2-\frac{2}{3}\vec{x}\cdot\vec{\theta}\Big).
\eenn
We can write the above using the property of a general POVM on $\mathbb{C}^2$, i.e. $\int_{\mathcal{X}}\D\mu(\vec{x})\vec{x}=0$ as
\benn
R_P^{sq}(\rho_\theta, \rho_B^{\pi^*})=\frac{1}{2}\Big\{\|\vec{\theta}\|^2+\frac{1}{9}\int_{\mathcal{X}}\D\mu(\vec{x})\|\vec{x}\|^2+\frac{1}{9}\int_{\mathcal{X}}\D\mu(\vec{x})\|\vec{x}\|^2\vec{x}\cdot\vec{\theta}-\frac{2}{3}\int_{\mathcal{X}}\D\mu(\vec{x})\sum_{j,k}\theta_j\theta_kx_j x_k\Big\}.
\eenn

\textit{Note: Although in reference \cite{Komaki2017} it is assumed  that the POVM P is rank-1, the above expressions hold in general for any POVM P, i.e. the vector $\vec{x}$ in \eqref{eq34} need not be a unit vector.}
\epr


\begin{lemma}\label{lem4}
For any POVM P, the average risk of the Bayes estimator with respect to the prior $\pi^*$ satisfies the inequalities:
\benn
\int_\Theta \txt{d}\pi^*(\theta)R_P^{rel}(\rho_\theta, \rho_B^{\pi^*})&\geq&\frac{1}{2}\log\frac{9}{2}-\frac{1}{6}\log2,\\
\int_\Theta \txt{d}\pi^*(\theta)R_P^{sq}(\rho_\theta, \rho_B^{\pi^*})&\geq&\frac{2}{9},
\eenn
for relative entropy and Hilbert-Schmidt distance respectively.
\bpr
(a) For~\textit{relative entropy}~:\\
From \cref{risklem} and the properties of the prior $\pi^*$ we get
\benn
\int_\Theta \txt{d}\pi^*(\theta)R_P^{rel}(\rho_\theta, \rho_B^{\pi^*})&=&\int_\Theta \txt{d}\pi^*(\theta)\Bigg\{-h\Bigg(\frac{1+\|\theta\|}{2}\Bigg)+\frac{1}{2}\log\frac{9}{2}-\frac{\log2}{2}\int_{\mathcal{X}}\D\mu(\vec{x})\sum\limits_{j,k}\theta_j\theta_k x_j x_k\Bigg\}\\
&=&\frac{1}{2}\log\frac{9}{2}-\frac{\log2}{2}\int_{\mathcal{X}}\D\mu(\vec{x})\sum\limits_{j,k}\frac{\delta_{jk}}{3} x_j x_k\\
&=&\frac{1}{2}\log\frac{9}{2}-\frac{1}{6}\log2\sum_j\mathbb{E}_\mu[r_j^2].
\eenn
However, since $\sum_j r_j^2\leq1$, it implies $\sum_j\mathbb{E}_\mu[r_j^2]\leq1$, and we obtain the required inequality :
\benn
\int_\Theta \txt{d}\pi^*(\theta)R_P^{rel}(\rho_\theta, \rho_B^{\pi^*})\geq \frac{1}{2}\log\frac{9}{2}-\frac{1}{6}\log2.
\eenn

(b) For \textit{Hilbert-Schmidt distance} :\\
From \cref{risklem} and the properties of the prior $\pi^*$ we get
\begin{flalign*}
\begin{split}
\int_\Theta \txt{d}\pi^*(\theta)R_P^{f_2}(\rho_\theta, \rho_B^{\pi^*})&=\int_\Theta \txt{d}\pi^*(\theta)\frac{1}{2}\Big\{\|\vec{\theta}\|^2+\frac{1}{9}\int_{\mathcal{X}}\D\mu(\vec{x})\|\vec{x}\|^2+\frac{1}{9}\int_{\mathcal{X}}\D\mu(\vec{x})\|\vec{x}\|^2\vec{x}\cdot\vec{\theta}-\\
&\frac{2}{3}\int_{\mathcal{X}}\D\mu(\vec{x})\sum_{j,k}\theta_j\theta_kx_j x_k\Big\}\\
&=\frac{1}{2}\Big\{1+\frac{1}{9}\sum_j\mathbb{E}_\mu[r_j^2]-\frac{2}{9}\sum_j\mathbb{E}_\mu[r_j^2]\Big\}\\
&=\frac{1}{2}\Big\{1-\frac{1}{9}\sum_j\mathbb{E}_\mu[r_j^2]\Big\}.
\end{split}
\end{flalign*}
Again, as $\sum_j r_j^2\leq1$, it implies $\sum_j\mathbb{E}_\mu[r_j^2]\leq1$, and we obtain the required inequality :
\benn
\int_\Theta \txt{d}\pi^*(\theta)R_P^{f_2}(\rho_\theta, \rho_B^{\pi^*})\geq \frac{1}{2}\big(1-\frac{1}{9}\big)=\frac{4}{9}.
\eenn
\epr
\end{lemma}


\begin{lemma}\label{lem5}
For any POVM $P^*$ that satisfies $\mathbb{E}_\mu[r_ir_j]=\frac{1}{3}\delta_{ij}$, the average risk of the Bayes estimator coincides with the  worst-case risk:
\benn
\int_\Theta \txt{d}\pi^*(\theta)R_{P^*}(\rho_\theta, \rho_B^{\pi^*})=\sup_\theta R_{P^*}(\rho_\theta, \rho_B^{\pi^*}).
\eenn
\bpr (i) (a) For \textit{relative entropy} : (a) See \cite[pg.~11]{Komaki2017}.\\

(b) For \textit{Hilbert-Schmidt} distance:\\

As $\mathbb{E}_\mu[r_ir_j]=\frac{1}{3}\delta_{ij}$, it implies $\sum_j\mathbb{E}_\mu[r_j^2]=1$, and thus by \cref{lem4}, 
\benn
\int_\Theta \txt{d}\pi^*(\theta)R_{P^*}^{f_2}(\rho_\theta, \rho_B^{\pi^*})=\frac{4}{9}.
\eenn
Now, for such a POVM $P^*$, $\|\vec{x}\|^2=1$, and the risk, see \cref{risklem}, becomes
\benn
R_{P^*}^{sq}(\rho_\theta, \rho_B^{\pi^*})&=&\frac{1}{2}\Big(\|\vec{\theta}\|^2+\frac{1}{9}-\frac{2}{9}\|\vec{\theta}\|^2\Big).
\eenn
Clearly, 
\benn
\max_{\|\theta\|} R_{P^*}^{sq}(\rho_\theta, \rho_B^{\pi^*})=\frac{4}{9}, \txt{at $\|\vec{\theta}\|=1$.}
\eenn
This implies that
\benn
\int_\Theta \txt{d}\pi^*(\theta)R_{P^*}(\rho_\theta, \rho_B^{\pi^*})=\sup_\theta R_{P^*}(\rho_\theta, \rho_B^{\pi^*}),~\txt{for both relative entropy and squared-distance.}
\eenn
\epr
\end{lemma}
\begin{restatable}{lemma}{qubitlemma}\label{lem1}
The uniform Haar measure on $\mathbb{S}^2$ is a least favourable prior for spherical 2-designs in $\mathbb{C}^2$.
\end{restatable}
\bpr
Firstly, note that a POVM $P^*$ with $\mathbb{E}_\mu[r_ir_j]=\frac{1}{3}\delta_{ij}$ is a spherical 2-design. (See \cref{def3} of spherical t-designs. Examples of spherical 2-designs include the SIC-POVM~\cite{Scott&Blume-Kohout2004} on $\mathbb{C}^2$ as well as the POVM defined through the Pauli measurements.) It can be seen from \cref{risklem} that the risk is a polynomial function of degree 2 in the variables $x,y,z$. It is straightforward to see that the average of the typical term $x_ix_j$ with respect to the Haar measure on $\mathbb{S}^2$ is $\frac{\delta_{i,j}}{3}$. Thus, any POVM with $\mathbb{E}_\mu[x_ix_j]=\frac{1}{3}\delta_{ij}$ is a spherical 2-design. Let us now proceed with the proof of the lemma. As the Bayes estimator $\hat{\rho}_B^{\pi}$ minimizes the average risk with respect to the prior $\pi$, 
\benn
\sup_\theta R_{P^*}(\rho_\theta, \rho_B^{\pi^*})=\sup_\pi\int_\Theta \txt{d}\pi(\theta)R_{P^*}(\rho_\theta, \rho_B^{\pi^*})\geq\sup_\pi\int_\Theta\txt{d}\pi(\theta)R_{P^*}(\rho_\theta, \rho_B^{\pi}).
\eenn
However, we just proved in \cref{lem5} that
\benn
\sup_\theta R_{P^*}(\rho_\theta, \rho_B^{\pi^*})&=&\int_\Theta \txt{d}\pi^*(\theta)R_{P^*}(\rho_\theta, \rho_B^{\pi^*}),\\
\implies \int_\Theta \txt{d}\pi^*(\theta)R_{P^*}(\rho_\theta, \rho_B^{\pi^*})&\geq&\sup_\pi\int_\Theta\txt{d}\pi(\theta)R_{P^*}(\rho_\theta, \rho_B^{\pi}).
\eenn
The other direction of the above inequality holds trivially, i.e.
\benn
\int_\Theta \txt{d}\pi^*(\theta)R_{P^*}(\rho_\theta, \rho_B^{\pi^*})\leq \sup_\pi\int_\Theta\txt{d}\pi(\theta)R_{P^*}(\rho_\theta, \rho_B^{\pi}),\\
\implies  \int_\Theta \txt{d}\pi^*(\theta)R_{P^*}(\rho_\theta, \rho_B^{\pi^*})=\sup_\pi\int_\Theta\txt{d}\pi(\theta)R_{P^*}(\rho_\theta, \rho_B^{\pi}).
\eenn
Thus, $\pi^*$ is a least favourable prior for a spherical 2-design in $\mathbb{C}^2$.
\epr
\begin{theorem}\label{thm4}
Any spherical 2-design for $\mathbb{C}^2$ is a minimax POVM.
\end{theorem}

\bpr
Recalling \cref{def3} of a minimax POVM and the fact that the Bayes estimator minimizes the average risk, we have:
\benn
\inf_P \inf_{\hat{\rho}}\sup_\theta R_P(\rho_\theta,\hat{\rho})&=&\inf_P \inf_{\hat{\rho}}\sup_\pi\int_\Theta \txt{d}\pi(\theta)R_P(\rho_\theta,\hat{\rho})\\
&\geq& \inf_P \sup_\pi\int_\Theta \txt{d}\pi(\theta)R_P(\rho_\theta,\rho_B^\pi)\\
&\geq&\inf_P \int_\Theta \txt{d}\pi^*(\theta)R_P(\rho_\theta,\rho_B^{\pi^*}).
\eenn
Now, \cref{lem4} and \cref{lem5} together imply that the average risk with respect to $\pi^*$ is minimized by the POVM $P^*$, i.e.
\benn
\inf_P\int_\Theta \txt{d}\pi^*(\theta)R_P(\rho_\theta, \rho_B^{\pi^*})=\int_\Theta \txt{d}\pi^*(\theta)R_{P^*}(\rho_\theta, \rho_B^{\pi^*}).
\eenn
Also, by \cref{lem1}, $\pi^*$ is a least favourable prior which means that $\rho_B^{\pi^*}$ is a minimax estimator. Therefore, we have
\benn
\int_\Theta \txt{d}\pi^*(\theta)R_{P^*}(\rho_\theta, \rho_B^{\pi^*})=\sup_\theta R_{P^*}(\rho_\theta, \rho_B^{\pi^*})=\inf_{\hat{\rho}}\sup_\theta R_{P^*}(\rho_\theta, \hat{\rho}).
\eenn
Thus, we obtain
\benn
\inf_P \inf_{\hat{\rho}}\sup_\theta R_P(\rho_\theta,\hat{\rho})\geq\inf_{\hat{\rho}}\sup_\theta R_{P^*}(\rho_\theta, \hat{\rho}).
\eenn
The other direction of the inequality above holds trivially, i.e.
\benn
\inf_P \inf_{\hat{\rho}}\sup_\theta R_P(\rho_\theta,\hat{\rho})\leq\inf_{\hat{\rho}}\sup_\theta R_{P^*}(\rho_\theta, \hat{\rho})
\eenn
Hence, we have proved that $P^*$, a spherical 2-design is a minimax POVM, i.e. 
\benn
\inf_P \inf_{\hat{\rho}}\sup_\theta R_P(\rho_\theta,\hat{\rho})=\inf_{\hat{\rho}}\sup_\theta R_{P^*}(\rho_\theta, \hat{\rho}).
\eenn
\epr
\section{Discussion \& Future work}\label{sec5}
To summarize, we extended the work done in reference \cite{Komaki2017} on minimax analysis to Bregman divergences. Moreover, by re-formulating Holevo's theorem~\cite[pg.~171]{holevo82} for the \textit{covariant state estimation problem} in terms of estimators, we found that a \textit{covariant} POVM is, in fact, minimax with Bregman divergence as the distance-measure. In addition to that, we found that it suffices that a measurement be covariant only under a subgroup $H$ of $G$ such that the unitary representation of $H$ forms a unitary 2-design for it to be minimax. Finally, in order to understand the problem of finding a minimax POVM for an arbitrary quantum state, we studied the problem for a qubit observing that a spherical 2-design defines a minimax POVM for a qubit.

In the covariant state estimation problem, we assume that the underlying group $G$ is compact. It is natural to ask if these results can be extended to infinite-dimensional systems, or equivalently, non-compact groups. The natural system that comes to mind when one thinks of an infinite-dimensional system is the set of coherent states of a Harmonic oscillator. The underlying group is the translation group $\mathcal{T}$ acting on the complex plane. The projective unitary representation of which is the Weyl-Heisenberg translation operator $\{D(\alpha)~|~\alpha\in\mathbb{C}\}$. Now, the translation group is non-compact. This means that one cannot define a normalisable measure on the group. Our derivation of the main result  on covariant state estimation, \cref{thm3*}, to obtain a minimax measurement uses a Bayesian approach. Recall that a minimax measurement is the one that minimizes the worst-case risk of a minimax estimator. Thus, we are interested in the following expression~: 
\begin{equation*}
    \inf_P \inf_{\hat{\rho}}\sup_\theta R_P(\rho_\theta,\hat{\rho}).
\end{equation*}{
The very first step of the proof involves re-writing the supremum over $\theta$ as a supremum over the probability distributions on $\Theta$, i.e. 
\begin{equation*}
   \inf_P \inf_{\hat{\rho}}\sup_\theta R_P(\rho_\theta,\hat{\rho})=\inf_P \inf_{\hat{\rho}}\sup_\pi \int_\Theta \D\pi(\theta) R_P(\rho_\theta,\hat{\rho}). 
\end{equation*}
Obviously, we cannot do so in the case of the translation group $\mathcal{T}$ that acts on the complex plane. So, our approach will not apply to the most general problem of estimating coherent states generated by the Weyl-Heisenberg translation operator $\{D(\alpha)~|~\alpha\in\mathbb{C}\}$. Indeed, a more general theorem for the case of locally compact groups \cite{bogomolov1982minimax, Hayashi2017} shows that covariant measurements minimize the worst-case risk (average risk cannot be defined for non-compact groups). However, the formalism considered in \cite{bogomolov1982minimax} does not include \emph{estimators}. It would be interesting to extend the same to our setting and, moreover, to come up with an appropriate definition of a Bayesian estimator for such cases.

The next obvious extension of this work is to find minimax POVMs for an arbitrary quantum state. It would be interesting to see if some kind of a t-design comes out as a solution. However, this requires a more generalized approach than mere brute-force calculations which become tedious in higher dimensions. Moreover, one could also generalize this result to arbitrary distance-measures such as Fidelity and Renyi divergences. The authors of reference \cite{ferrie&kohout} have derived the Bayes estimator for distance-measures based on Bhattacharya distance. Partial results \cite{Keung&Ferrie} are known for fidelity as the distance-measure, but the Bayes estimator remains unknown for a general state with fidelity as the distance-measure. But, these generalizations are not so straight forward either and require a different technique. 
\begin{acknowledgements}
We acknowledge that reference \cite{Hayashi2017} was brought to our attention by an anonymous reviewer. CF and MT acknowledge Australian Research Council Discovery Early Career Researcher Awards, projects No. DE170100421 and DE160100821, respectively.
\end{acknowledgements}
\bibliographystyle{apsrev4-1}
\bibliography{draft}
\begin{appendix}
\section{Quantum Bayes estimator for Bregman divergence}\label{A0}
\begin{theorem}\label{thm1g}If the loss function is Bregman divergence, see \cref{def2}, then
\benn
\mathbb{E}_{\theta}\mathbb{E}_{X|\theta}[D_f(\rho_{\theta},\hat{\rho}(X))-D_f(\rho_{\theta},\hat{\rho}_B(X))] \geq 0,
\eenn
for all states $\theta\in\Theta$ and estimators $\hat{\rho}$, where $\hat{\rho}_B$ is the Bayes estimator, see Equation\eqref{eq3}.
\end{theorem}
\bpr
\benn
\begin{split}
\mathbb{E}_{\theta}\mathbb{E}_{X|\theta}[D_f(\rho_{\theta},\hat{\rho}(X))-D_f(\rho_{\theta},\hat{\rho}_B(X))]=&\int_\Theta \D\pi(\theta)\int_\mathcal{X}\D p(x|\theta)\Tr\big[f(\rho_\theta)-f(\hat{\rho}(x))-\\
&f'(\hat{\rho}(x))(\rho_\theta-\hat{\rho}(x))-f(\rho_\theta)+f(\hat{\rho}_B(x))+f'(\hat{\rho}_B(x))(\rho_\theta-\hat{\rho}_B(x))\big]\\
=&\int_\Theta \D\pi(\theta)\int_\mathcal{X}\D p(x|\theta)\Tr\big[f(\hat{\rho}_B(x))-f(\hat{\rho}(x))-f'(\hat{\rho}(x))(\rho_\theta-\hat{\rho}(x))+\\
&f'(\hat{\rho}_B(x))(\rho_\theta-\hat{\rho}_B(x))\big]\\
=&\int_\mathcal{X}\D p_\pi(x)\Tr\big[f(\hat{\rho}_B(x))-f(\hat{\rho}(x))-f'(\hat{\rho}(x))(\hat{\rho}_B(x)-\hat{\rho}(x))\big]+\\
&\int_\mathcal{X}\D p_\pi(x)\Tr\big[f'(\hat{\rho}_B(x))(\hat{\rho}_B(x)-\hat{\rho}_B(x))\big]\\
=&\int_\mathcal{X}\D p_\pi(x)D_f(\hat{\rho}_B(x),\hat{\rho}(x))\geq 0.
\end{split}
\eenn
The last inequality follows from the non-negativity of Bregman divergence.
\epr
\begin{corollary}
For all \textit{a-priori} probability distributions $\pi_{\Theta}(\theta)$ over the parameter space $\Omega_{\Theta}$,
\benn
r(\pi,\hat{\rho})\geq r(\pi,\hat{\rho}_B),
\eenn
i.e. the Bayes estimator minimizes the average risk for Bregman divergence.
\end{corollary}

\section{Proof of \cref{lem0}}\label{A1}
We first present the Radon-Nikodym theorem for operator-valued measures \cite{Radon-Nikodym}, without proof, as stated in~~\cite[pg.~167]{holevo82}.
\begin{prop}[Radon-Nikodym theorem for operator-valued measures]
Let $(\mathcal{X},\Sigma)$ be a measurable space and let $\{M(B);~B\in\Sigma\}$ be an additive operator-valued function dominated by a measure $\{m(B);~B\in\Sigma\}$ in the sense that
\benn
|\langle\phi|M(B)|\psi\rangle|\leq m(B)\|\phi\|~\|\psi\|, ~~~~B\in\Sigma,
\eenn
for all $\phi,~\psi\in\mathcal{H}$. Then, there exists an operator-valued function P(.) defined uniquely for m-almost all $ x\in\mathcal{X}$ (i.e. for all x except for a set of zero $m$-measure), satisfying $\|P(x)\|\leq1$ such that 
\benn
\langle\phi|M(B)|\psi\rangle=\int_B \langle \phi|P(x)|\psi\rangle m(\txt{d}x), ~~~~B\in\Sigma
\eenn
for all $\phi,~\psi\in\mathcal{H}$. If $M(B)\geq 0$ for all $B\in\Sigma$, then $P(x)\geq 0$ for m-almost all $x\in\mathcal{X}$.
\end{prop}

\RadonNikodym*
\bpr 
Define $\mu(B)=\Tr [P(B)]$, then 
\benn
\langle \phi|P(B)|\phi\rangle \leq \mu(B),~~~~\forall~|\phi\rangle\in\mathcal{H}.
\eenn
By Cauchy-Schwarz inequality, 
\benn
|\langle\phi| P(B)|\psi\rangle|\leq \mu(B),~~~~\forall |\phi\rangle,~|\psi\rangle\in\mathcal{H}.
\eenn
Thus, P(B) is dominated by $\mu(B)$. By the Radon-Nikodym theorem, it admits a density M(x) defined uniquely $\mu-$ almost everywhere:
\benn
P(B)=\int_B \mu(dx) M(x).
\eenn
As $P(B)\geq0, M(x)\geq 0$. Taking trace on both sides of the above equation implies $\Tr[M(x)]=d$ $\mu-$ almost everywhere.
\epr

\section{Lower semi-continuity of Bregman divergence}\label{lsc}
Before proving the lower semi-continuity of $D_f$, we state the following lemma which will be used in the proof.
\begin{lemma}\label{lemLSC}
The map $\rho\mapsto \Tr f(\rho)$ is continuous on $\Omega=[0,1]$.
\end{lemma}
\bpr 
Consider a sequence of density operators $(\rho_n)_n$ that weakly converge to a density operator $\rho$, i.e. $\Tr \rho_n a\rightarrow \Tr\rho a$ for all $a\in\mathcal{B}(\mathcal{H})$, where $\mathcal{H}$ is a finite dimensional Hilbert space. By choosing $[a]=|i\X j|$, where $|i\X j|$ is a basis in $\mathcal{B}(\mathcal{H})$, we obtain element-wise convergence of $\rho_n$ to $\rho$, which in turn implies the convergence of the corresponding eigenvalues. 

Now, $\Tr f(\rho_n)=\sum_{j=1}^{rk(\rho_n)} f(\lambda_j^n)$ where $\{\lambda_j^n\}_{j=1}^{rk(\rho_n)}$ are the eigenvalues of $\rho_n$. But, as f is continuous on [0,1], $f(\lambda_j^n)\rightarrow f(\lambda_j)$, where $\{\lambda_j\}_{j=1}^{rk(\rho)}$ are the eigenvalues of $\rho$. Thus, the sum $\Tr f(\rho_n)=\sum_{j=1}^{rk(\rho_n)} f(\lambda_j^n)$ also converges to $\Tr f(\rho)=\sum_{j=1}^{rk(\rho)} f(\lambda_j)$, and this proves the continuity of $\rho\mapsto \Tr f(\rho)$ on [0,1].
\epr

\begin{theorem}
Bregman divergence $D_f(.,.)$ is lower semi-continuous.
\end{theorem}
\bpr 
We generalize the proof of lower semi-continuity of relative entropy as given in reference \cite{Wehrl}. To begin with, we show that there exists a representation of $d_f$ in which the argument of trace does not contain a product of two non-commuting operators. In order to do so, let us define a quantity $D_f^\lambda$ as
\benn
D_f^\lambda (\rho,\sigma)=\frac{1}{\lambda}\Tr\big(\lambda f (\rho) + (1-\lambda)f(\sigma)-f(\lambda\rho+(1-\lambda)\sigma)\big),
\eenn
where $\lambda\in(0,1)$. It is straightforward to verify that the map $\lambda\mapsto\lambda D_f^\lambda$ is concave on the interval $(0,1)$. Note that $\frac{\D}{\D\lambda}(\lambda D_f^\lambda)\big|_{\lambda=0}=D_f$. Thus, as the map $\lambda\mapsto\lambda D_f^\lambda$ is concave on (0,1), the slope at $\lambda=0$ will always be greater than the differences $\frac{\lambda D_f^\lambda-0.D_f^\lambda}{\lambda}$ for all $\lambda\in(0,1)$. Assuming that $0.D_f^0=0$, we have
\benn
\lambda D_f^\lambda \leq \lambda D_f
\Leftrightarrow D_f^\lambda \leq D_f,  ~~~~\forall\lambda\in(0,1).
\eenn
As $\lim\limits_{\lambda\rightarrow 0}D_f^\lambda=D_f$, we have
\benn
\sup_\lambda D_f^\lambda (\rho,\sigma)=D_f(\rho,\sigma).
\eenn

Let $\rho_n\Rightarrow\rho$ and $\sigma_n\Rightarrow\sigma$ be given. Then,  by \cref{lemLSC}, the map $(\rho,\sigma)\mapsto D_f^\lambda(\rho,\sigma)$ is continuous. Therefore,
\benn
D_f(\rho,\sigma)&=&\sup_\lambda D_f^\lambda(\rho,\sigma)\\
&=&\sup_\lambda \lim_{n\rightarrow\infty} D_f^\lambda(\rho_n,\sigma_n)\\
&\leq&\liminf_{n\rightarrow\infty}\sup_\lambda D_f^\lambda(\rho_n,\sigma_n)\\
&=&\liminf_{n\rightarrow\infty} D_f(\rho_n,\sigma_n).
\eenn
But, $D_f(\rho,\sigma)\leq\liminf\limits_{n\rightarrow\infty} D_f(\rho_n,\sigma_n)$ defines a lower semi-continuous function.
\epr

\section{Why the Bayes estimator is discontinuous.}\label{disctsBayes}
For the Bayes estimator to be continuous in the prior, it should hold that for any convergent sequence $(\pi_n)_n$, 
\[
\lim_{n\rightarrow\infty}\hat{\rho}_B^{\pi_n}(x)=\hat{\rho}_B^{\lim\limits_{n\rightarrow\infty}\pi_n}(x), ~~~~\forall x\in\mathcal{X}.
\]
Below is an example that shows that the above is \emph{not} true in general.
\begin{eg}
Let us assume that we are doing a $\sigma_z$ measurement, thus, $\mathcal{X}=\{0,1\}$. Now, consider a sequence of priors $(\pi_n)_n$ that converges to $\mu(\theta)=\delta(\theta-\theta_0)$ where $\theta_0$ corresponds to $|0\X0|$ and let each element of the sequence be defined as below:
\[
\pi_n(\theta)=\big(1-\frac1n\big)\delta(\theta-{\theta_0})+\frac1n\delta(\theta-{\theta_1}),
\]
with $\theta_1$ corresponding to $|1\X1|$. Then, the Bayes estimator, see Equation \eqref{eq13}, for each element of the sequence is
\benn
\hat{\rho}_B^{\pi_n}(x)
&=&\big(1-\frac1n\big)\frac{ p(x|\theta_0)\rho_{\theta_0}}{(1-\frac1n)p(x|\theta_0)+\frac1np(x|\theta_1)}+\frac1n\frac{ p(x|\theta_1)\rho_{\theta_1}}{(1-\frac1n)p(x|\theta_0)+\frac1np(x|\theta_1)},
\eenn
which in the limiting case reduces to
\[\lim_{n\rightarrow\infty}\hat{\rho}_B^{\pi_n}(x)=\begin{cases}
							\rho_{\theta_0} &\quad \txt{if x=0},\\
                            \rho_{\theta_1} &\quad \txt{if x=1}.
						  \end{cases}
\]
But, the Bayes estimator for the limit $\mu$ of the sequence $(\pi_n)_n$ is
\[
\hat{\rho}_B^{\mu}(x)=\begin{cases}
						\rho_{\theta_0} &\quad \txt{if x=0},\\
						\txt{not defined} &\quad \txt{if x=1}.
					    \end{cases}	
\]
Since the Bayes estimator for $\mu$ is not defined at $x=1$, we can define it to be 
\benn
\hat{\rho}_B^{\mu}(x=1)=\lim_{n\rightarrow\infty}\hat{\rho}_B^{\pi_n}(x=1).
\eenn
However, for the Bayes estimator to be continuous in the prior the above should be true for all sequences of priors that have the same limit point. Let us consider another sequence $(\mu_n)_n$ that converges to $\mu$ with each element defined as below:
\[
\mu_n(\theta)=\big(1-\frac1n\big)\delta(\theta-{\theta_0})+\frac1n\delta(\theta-{\theta_+}),
\]
where $\theta_+$ corresponds to $|+\X+|$. Then, the Bayes estimator for $\mu_n$ would be 
\benn
\hat{\rho}_B^{\mu_n}(x)=\big(1-\frac1n\big)\frac{ p(x|\theta_0)\rho_{\theta_0}}{(1-\frac1n)p(x|\theta_0)+\frac1np(x|\theta_+)}+\frac1n\frac{ p(x|\theta_+)\rho_{\theta_+}}{(1-\frac1n)p(x|\theta_0)+\frac1np(x|\theta_+)},
\eenn
which in the limiting case reduces to 
\[\lim\limits_{n\rightarrow\infty}\hat{\rho}_B^{\mu_n}(x)=\begin{cases}
						  \rho_{\theta_0} &\quad \txt{if x=0},\\
                          \rho_{\theta_+} &\quad \txt{if x=1}.
						 \end{cases}
\] 
Now, if we again chose to define the Bayes estimator for $\mu$ at $x=1$ as $\lim\limits_{n\rightarrow\infty}\hat{\rho}_B^{\mu_n}(x=1)$, we would run into a contradiction since $\lim\limits_{n\rightarrow\infty}\hat{\rho}_B^{\mu_n}(x=1)\neq \lim\limits_{n\rightarrow\infty}\hat{\rho}_B^{\pi_n}(x=1)$.
\end{eg}

The above example establishes that the Bayes estimator does not admit a continuous extension on the null set (where it is not defined) of the given prior.

\section{ Additional lemma(s)}
\begin{lemma}\label{lemC1}
Given a self-adjoint operator $A$ parametrized by u, the derivative of a function of the operator with respect to the parameter at $u=u_0$ is given by
\benn
\frac{d}{du}\Tr\Big[f\Big(A(u)\Big)\Big]\Bigg|_{u=u_0}=\Tr\Big[A'(u)\Big|_{u=u_0}f'\Big(A(u_0)\Big)\Big]
\eenn
\bpr
See reference \cite{Tsallis}.
\epr
\end{lemma}
\begin{lemma}\label{lemE2}
Consider a continuous function $\mathcal{F}$ defined on a compact set $\mathscr{B}$ and closed subsets $B_x\subseteq \mathscr{B}$ such that $B_1 \subseteq B_2 ...\subseteq \mathscr{B}$. Then, assuming that the sequence $(B_x)_x$ is dense in $\mathscr{B}$, the following holds:
\benn
\lim_{x\rightarrow\infty} \sup_{B_x}\mathcal{F}=\sup_{\mathscr{B}}\mathcal{F}.
\eenn
\end{lemma}
\bpr
As $B_1 \subseteq B_2 ...\mathscr{B}$, it implies that
\be\label{eq42}
\lim_{x\rightarrow\infty} \sup_{B_x}\mathcal{F}\leq\sup_{\mathscr{B}}\mathcal{F}.
\ee
Note that $\mathscr{B}$ is a compact set and hence
\benn
\sup_{\mathscr{B}}\mathcal{F}=\max_{\mathscr{B}}\mathcal{F}.
\eenn
Let $b:=\argmax_{\mathscr{B}}\mathcal{F}$. Then, as the sequence of subsets $(B_x)_x$ is dense in $\mathscr{B}$, it implies that there exists a sequence $(b_x)_x$ that converges to $b$ such that $b_x\in B_x$. Moreover, we know that
\benn
\sup_{B_x}\mathcal{F}\geq F(b_x).
\eenn
Thus, 
\benn
\lim_{x\rightarrow\infty}\sup_{B_x}\mathcal{F}\geq \lim_{x\rightarrow\infty}F(b_x).
\eenn
As $\mathcal{F}$ is continuous, we get
\be\label{eq43}
\lim_{x\rightarrow\infty}\sup_{B_x}\mathcal{F}\geq \mathcal{F}\big(\lim_{x\rightarrow\infty}b_x\big)=\mathcal{F}(b)=\sup_{\mathscr{B}}\mathcal{F}.
\ee
Equations \eqref{eq42} and \eqref{eq43} imply the result.
\epr
\begin{lemma}[\cite{holevo82}, Theorem 2.1]\label{lemE3}
Let $P_0$ be a positive operator in the representation space such that $[P_0,V_g]=0~\forall g \in G_0$, where $G_0$ is the stationary subgroup of G, and satisfying:
\be\label{lemE328}
\int_G V_gP_0V_g^{\dagger}\D\mu(g)=\mathbb{I}
\ee
Then setting $P(g\theta_0)=V_gP_0V_g^{\dagger}$, we get an operator-valued function of $\theta$ such that:
\be\label{lemE329}
M(B)=\int_B P(\theta)\D\nu(\theta),~~~~B\in\mathcal{A}(\Theta)
\ee
is a covariant measurement with respect to $g\mapsto V_g$. Conversely, for any covariant measurement $M(d\theta)$ there is a unique operator $P_0$ satisfying \eqref{lemE328} such that $M(B)$ can be expressed as in \eqref{lemE329}. $P_0$ is referred to as the \emph{seed} of the covariant measurement.
\end{lemma}

\section{Aside on t-designs}\label{A2}
\begin{definition}[Unitary t-design]\label{unitarytdesign}
Consider the set of unitary matrices U(d) on a d-dimensional Hilbert space $\mathcal{H}$. A unitary t-design is a finite subset $\{U_i\}_{i=1}^N\subset U(d)$, such that for all states $\rho\in\mathcal{S}(\mathcal{H})$ the following holds:
\benn
\frac{1}{N}\sum_{i=1}^N U_i^{\otimes t}\rho (U_i^\dagger)^{\otimes t}=\int_{U(d)}\D U  U^{\otimes t}\rho (U^\dagger)^{\otimes t},
\eenn
where `dU' is the Haar measure on U(d).
\end{definition}
\begin{definition}[Spherical t-design]\label{def3}
Consider the unit sphere $\mathbb{S}^{d-1}$ in the d-dimensional Euclidean space $\mathbb{R}^d$. A spherical t-design is a finite subset $\mathcal{S}\subset \mathbb{S}^{d-1}$, such that the average value of a polynomial f of degree $\leq t$ on $\mathcal{S}$ equals its average on $\mathbb{S}^{d-1}$:
\benn
\frac{1}{|\mathcal{S}|}\sum_{s_n\in\mathcal{S}} f(s_n)=\int_{\mathbb{S}^{d-1}}ds~f(s),
\eenn
where `ds' is the Lebesgue measure on $\mathbb{S}^{d-1}$.
\end{definition}


\end{appendix}

\end{document}